\documentclass[reprint,prd,nofootinbib]{revtex4}
\usepackage{amssymb}

\usepackage{amsmath}
\usepackage{graphicx}
\usepackage{dcolumn}
\usepackage{bm}
\usepackage{float}
\usepackage{enumitem}
\usepackage{hyperref}

\begin{document}

\title{Linear stability of Israel-Stewart theory in the presence of net-charge diffusion}
\author{C.\ V.\ Brito}
\email{caio\_brito@id.uff.br}
\author{G.\ S.\ Denicol}
\email{gsdenicol@id.uff.br}
\affiliation{Instituto de F\'{\i}sica, Universidade Federal Fluminense, UFF, Niter\'oi, 24210-346, RJ, Brazil}

\begin{abstract}
In this paper, we perform a linear stability analysis of Israel-Stewart theory around a global equilibrium state, including the effects of shear-stress tensor, net-baryon diffusion current and diffusion-viscous coupling. We find all the relevant modes of this theory and derive necessary conditions that these modes must satisfy in order to be stable and subluminal. With these conditions, we then derive constraints for the shear and diffusion relaxation times and the transport coefficients related to diffusion-viscous coupling. 
\end{abstract}

\maketitle

\section{\label{sec:1}Introduction}

Relativistic fluid dynamics is an effective theory constructed to
describe the long-wavelength and long-time dynamics of a many-body system, with applications that range from astrophysics \cite{Rezzollabook} to high-energy nuclear physics \cite{Hatsudabook}. In the past 20 years, there has been a growing interest in understanding fundamental aspects of relativistic hydrodynamics \cite{flork}, mostly due to the fluid-dynamical modeling of the quark-gluon plasma, produced in modern particle accelerators via ultrarelativistic heavy-ion collisions \cite{heinz, flork,kodama,gale}.

Relativistic generalizations of Navier-Stokes theory derived by Eckart \cite{eckart} and later by Landau and Lifshitz, independently, \cite{landaubook}, are known to be ill-defined, containing intrinsic instabilities when perturbed around an arbitrary global equilibrium state \cite{hw1,denicol2,rischke}. In Refs.~\cite{rischke, hiscock02}, such linear instabilities were shown to be related to the acausal nature of such theories, which allows perturbations to propagate with an infinite speed. These fundamental problems prohibit the application of traditional Navier-Stokes theory to describe any practical fluid-dynamical problem, may it be in the description of neutron star mergers or in the description of the quark-gluon plasma produced in heavy-ion collisions. 

\textit{Linearly} stable and causal theories of relativistic fluid dynamics were later derived by Israel and Stewart, following the procedure initially developed by Grad \cite{Grad} for nonrelativistic systems. Israel and Stewart performed this task in two distinct ways: the first being a phenomenological derivation based on the second law of thermodynamics \cite{is1}, and the second being a microscopic derivation starting from the relativistic Boltzmann equation \cite{is2}. Similar theories have been widely developed in the past decades \cite{MullerReviw, mueller, mueller2, carter, OG, BRSSS, denicol4, molnar, denicol5,peralta,liu}, but all carry the same fundamental aspects: In contrast to Navier-Stokes theory, such causal theories of fluid dynamics include in their description the transient dynamics of the nonconserved dissipative currents. For this reason, they were initially named by Israel and Stewart as transient fluid dynamics (nowadays, they are often referred to as \textit{second-order theories}). Here, we also note that novel causal extensions of \textit{first-order theories} have been recently presented. In this case, the fluid-dynamical equations can be rendered causal by the inclusion of first-order timelike gradients (and not only spacelike gradients, as traditionally done) \cite{bemfica, Andersson:2011fd} in the constitutive relations satisfied by the dissipative currents. The causality of such novel formulations of fluid dynamics is not guaranteed and was shown to depend on the matching conditions that define the local equilibrium state \cite{bemfica}. 

However, in this paper we focus our discussion on Israel-Stewart theory. At this point, it is important to remark that the theory formulated by Israel and Stewart is not guaranteed to be linearly causal and stable. As was first shown by Hiscock and Lindblom and, later, by Olson, such transient theories of fluid dynamics are only linearly causal and stable if their transport coefficients satisfy certain conditions \cite{hiscock02, olson}. Such conclusions were obtained by analyzing the properties of the theory in the linear regime and by imposing that the perturbations around a global equilibrium state are stable and propagate subluminally. More recent analyses were developed in Ref.~\cite{denicol2}, including only the effects of bulk viscosity, and, later, in Ref.~\cite{rischke}, which included the effects of both shear and bulk viscosity. In both these papers, constraints for the shear and bulk relaxation times were explicitly derived. Nowadays, causality analysis has even been performed also in the nonlinear regime \cite{jn1,jn2} (in this case, including the effects of shear and bulk viscosity), where more general inequalities required to ensure the causal propagation of the theory were derived. In the latter case, the inequalities constrain not only the transport coefficients, but also  the values of the dissipative currents (in the linear regime, the inequalities derived in Ref.~\cite{jn2} reduce to those derived in Refs.~\cite{denicol2,rischke}). Such constraints are relevant, e.g. for fluid-dynamical applications in heavy-ion collisions, since the transport coefficients of QCD matter are not precisely known (often, they are completely unknown) and constraints on transport coefficients (and the values of the dissipative currents) can be extremely useful.

Recently, several programs to experimentally study QCD matter at finite net-baryon density have been put in motion at the Relativistic Heavy Ion Collider, in Brookhaven National Lab (Upton, USA), and at the Nuclotron-based Ion Collider fAcility, in the Joint Institute for Nuclear Research (Dubna, Russia), and, will be starting soon at the Facility for Antiproton and Ion Research, in GSI Helmholtzzentrum f\"ur Schwerionenforschung (Darmstadt, Germany). Nevertheless, the more recent investigations on the stability and causality of fluid-dynamical descriptions \cite{jn1, jn2, denicol2, rischke} have not yet considered the \textit{complete} set of the Israel-Stewart equations, usually neglecting any dissipation by net-baryon diffusion and, also, possible diffusion-viscous coupling terms \footnote{The work by Olson \cite{olson} considered the complete Israel-Stewart equations, including all sources of fluctuations. Nevertheless, they did not consider the limit of vanishing background net charge and did not explicitly evaluate the dispersion relations for the hydrodynamic modes.}. In this paper, we actually perform a linear stability analysis around global equilibrium of Israel-Stewart theory, including the effects of the shear-stress tensor and net-baryon diffusion 4-current (all effects of bulk viscous pressure are neglected). We find all the relevant modes of this theory and derive the conditions that these modes must satisfy in order to be stable and subluminal. With this result, we obtain new conditions that the shear and diffusion relaxation times must satisfy so that Israel-Stewart theory remains linearly causal and stable. We further find constraints for the transport coefficients that couple the shear-stress tensor and the net-baryon diffusion current (diffusion-viscous coupling). In other words, we show that the inclusion of diffusion-viscous coupling in the equations of motion drives the theory unstable, if these transport coefficients do not satisfy certain bounds. Such novel constraints may be useful when such transport coefficients are included in the current fluid-dynamical simulations of the quark-gluon plasma.  

This paper is divided as follows. In Sec.~\ref{rel_hydro} we review the fundamentals of relativistic hydrodynamics, as proposed by Israel and Stewart \cite{is1}. Then, in Sec.~\ref{linear_hydro}, we linearize the Israel-Stewart equations around a global equilibrium state, expressing the resulting equations in Fourier space. In particular, we demonstrate how to decompose the linearized equations of motion in Fourier space into independent equations of motion for their longitudinal and transverse components -- a procedure that considerably simplifies the calculations. Next, Secs.~\ref{stab_wo} and \ref{stab_with} are dedicated to the study of the theory's linear stability in the absence and in the presence of diffusion-viscous coupling terms, respectively. All our conclusions are summarized in Sec.~\ref{conc}. In this paper, we use natural units $c=\hbar=k_{\mathrm{B}}=1$, and the \textit{mostly minus} convention for the metric tensor $g^{\mu\nu}=\mathrm{diag}(+,-,-,-)$.

\section{Relativistic Fluid Dynamics}
\label{rel_hydro}

In this section, we briefly review the equations of relativistic dissipative fluid dynamics. The main equations are the continuity equations associated with the conservation of net charge, energy, and momentum 
\begin{eqnarray}
\partial _{\mu }T^{\mu \nu } &=&0,  \label{tmunu} \\
\partial _{\mu }N^{\mu } &=&0,  \label{nmu}
\end{eqnarray}%
where $T^{\mu \nu }$ is the energy-momentum tensor and $N^{\mu }$ is the net-charge 4-current. In our case, the only conserved charge considered is the baryon number, with electric charge and strangeness being neglected.

It is convenient to write the conserved currents in a fluid-dynamical form,
\begin{eqnarray}
T^{\mu \nu } &=&\varepsilon u^{\mu }u^{\nu }-P\Delta ^{\mu \nu }+\pi ^{\mu
\nu }, \\
N^{\mu } &=&n_B u^{\mu }+n^{\mu },
\end{eqnarray}%
where $\varepsilon $ is the energy density, $n_B$ is the net-baryon number density, $P\left( n_B,\varepsilon \right) $ is the thermodynamic pressure, $u^{\mu }$ is the normalized 4-velocity ($u^{\mu}u_{\mu }=1$), $n^{\mu }$ is the net-baryon diffusion current, and $\pi
^{\mu \nu }$ is the shear-stress tensor. In this work, we employ the Landau-Lifshitz picture for the velocity field \cite{landaubook}, where it is defined as an eigenvector
of the energy-momentum tensor, i.e., $T^{\mu \nu }u_{\nu }\equiv \varepsilon
u^{\mu }$. We also introduced the projection operator onto the 3-space
orthogonal to $u^{\mu }$, $\Delta ^{\mu \nu }\equiv g^{\mu \nu }-u^{\mu
}u^{\nu }$. All effects of dissipation due to bulk viscous pressure are neglected in our calculations. Note that all the dissipative currents are constructed to be orthogonal to
the 4-velocity,%
\begin{equation}
u_{\mu }n^{\mu }=0,\left. {}\right. u_{\mu }\pi ^{\mu \nu }=0\text{.}
\label{ortho}
\end{equation}

When considering ideal fluids, the conservation laws supplemented by an equation of state are sufficient to describe the time evolution of the system. On the other hand, when considering viscous fluids, the conservation
laws must also be complemented by dynamical equations for the dissipative currents. Here, we employ relaxation-type equations derived from kinetic theory as our baseline
\cite{dnmr,denicol4,14moment}, 
\begin{align}
\tau _{n}\dot{n}^{\left\langle \mu \right\rangle }+n^{\mu }& =\kappa
_{n}\nabla ^{\mu }\alpha_B -n_{\nu }\omega ^{\nu \mu }-\delta _{nn}n^{\mu
}\theta +\ell _{n\pi }\Delta ^{\mu \nu }\nabla _{\lambda }\pi _{\nu
}^{\lambda }\text{ }  \notag \\
& -\tau _{n\pi }\pi ^{\mu \nu }\nabla _{\nu }P-\lambda _{nn}n_{\nu }\sigma
^{\mu \nu }-\lambda _{n\pi }\pi ^{\mu \nu }\nabla _{\nu }\alpha_B,
\label{eq_heat} \\
\tau _{\pi }\dot{\pi}^{\left\langle \mu \nu \right\rangle }+\pi ^{\mu \nu }&
=2\eta \sigma ^{\mu \nu }+2\tau _{\pi }\pi _{\lambda }^{\left\langle \mu
\right. }\omega ^{\left. \nu \right\rangle \lambda }-\delta _{\pi \pi }\pi
^{\mu \nu }\theta -\tau _{\pi \pi }\pi ^{\lambda \left\langle \mu \right.
}\sigma _{\lambda }^{\left. \nu \right\rangle }  \notag \\
& -\tau _{\pi n}n^{\left\langle \mu \right. }\nabla ^{\left. \nu
\right\rangle }P+\ell _{\pi n}\nabla ^{\left\langle \mu \right. }n^{\left.
\nu \right\rangle }+\lambda _{\pi n}n^{\left\langle \mu \right. }\nabla
^{\left. \nu \right\rangle }\alpha_B ,  \label{eq_shear}
\end{align}%
where $\alpha_B\equiv \mu_B /T$, with $\mu_B$ being the baryon chemical potential and $T$ the temperature. Above, we
defined the comoving time derivative, $\dot{A}\equiv u^{\mu }\partial _{\mu
}A$, the expansion rate, $\theta \equiv \partial _{\mu }u^{\mu }$, the shear
tensor $\sigma ^{\mu \nu }\equiv \partial ^{\left\langle \mu \right.
}u^{\left. \nu \right\rangle }$, the vorticity tensor, $\omega ^{\mu \nu
}\equiv \left( \nabla ^{\mu }u^{\nu }-\nabla ^{\nu }u^{\mu }\right) /2$, and
the projected derivative $\nabla ^{\mu }$ $\equiv $ $\partial ^{\left\langle
\mu \right\rangle }$. We further employ the notation, $A^{\left\langle \mu
\right\rangle }\equiv \Delta _{\nu }^{\mu }A^{\nu }$, and $A^{\left\langle
\mu \nu \right\rangle }\equiv \Delta _{\alpha \beta }^{\mu \nu }A^{\alpha
\beta }$, where we make use of the double, symmetric, and traceless
projection operator $\Delta _{\alpha \beta }^{\mu \nu }=\left( \Delta
_{\alpha }^{\mu }\Delta _{\beta }^{\nu }+\Delta _{\beta }^{\mu }\Delta
_{\alpha }^{\nu }\right) /2-\Delta ^{\mu \nu }\Delta _{\alpha \beta }/3$.

In the equations of motion for the dissipative currents, we introduced a wide set of transport coefficients, all being functions of temperature and
chemical potential. The most relevant for our work are the net-charge diffusion coefficient, $\kappa _{n}$, the shear viscosity coefficient, $\eta $, and the diffusion and shear relaxation times, $\tau _{n}$ and $\tau _{\pi }$, respectively. The effects of
the relaxation times have already been extensively investigated in the linear regime in \cite{hiscock02, olson, rischke} and were shown to be essential to render the theory causal and stable. So far, the remaining coefficients have not been widely
investigated, even in the linear regime. Such coefficients can be relevant as they determine the strength of the second-order terms that appear in the
fluid-dynamical equations. In particular, we are interested in the effects
of the coupling terms $\Delta ^{\mu \nu }\nabla _{\lambda }\pi _{\nu
}^{\lambda }$ and $\nabla ^{\left\langle \mu \right. }n^{\left. \nu
\right\rangle }$, which are associated with the transport coefficients $\ell
_{n\pi }$ and $\ell _{\pi n}$, respectively. The effect of these transport coefficients can be investigated in the linear regime. We note that such coupling terms also appear in the phenomenological derivation of Israel-Stewart theory \cite{is1, is2}.

For the sake of convenience, we further reexpress the conservation laws, Eqs.~(\ref{tmunu}) and (\ref{nmu}), in the following way,

\begin{align}
u_{\nu }\partial _{\mu }T^{\mu \nu }& =\dot{\varepsilon}+\,\left(
\varepsilon +P\right) \theta -\pi ^{\alpha \beta }\sigma _{\alpha \beta }=0,
\label{EoM1} \\
\Delta _{\nu }^{\lambda }\partial _{\mu }T^{\mu \nu }& =\left( \varepsilon
+P\right) \dot{u}^{\lambda }-\nabla ^{\lambda }P-\pi ^{\lambda \beta }\dot{u}%
_{\beta }+\Delta _{\nu }^{\lambda }\nabla _{\mu }\pi ^{\mu \nu }=0,
\label{EoM2} \\
\partial _{\mu }N^{\mu }& =\dot{n}_B+n_B \theta -n^{\mu }\dot{u}_{\mu }+\nabla
_{\mu }n^{\mu }=0.  \label{EoM3}
\end{align}%
The fluid-dynamical equations will be linearized in the above form.

\section{Linearized fluid dynamics}
\label{linear_hydro}

In this section, we linearize the fluid-dynamical equations described in the previous section around a global equilibrium state. For this purpose, we consider small fluid-dynamical perturbations around a hydrostatic equilibrium state, with an energy density, $\varepsilon _{0}$, a vanishing net-baryon number density, $n_{B,0}=0$, a constant 4-velocity,  $u_{0}^{\mu }$, and vanishing dissipative currents, $n_{0}^{\mu }=\pi _{0}^{\mu \nu }=0$.

Perturbations around such an equilibrium state can be expressed in the following form, 
\begin{gather}
\varepsilon =\varepsilon _{0}+\delta \varepsilon ,\left. {}\right.
n_{B}=\delta n_{B},\left. {}\right. u^{\mu }=u_{0}^{\mu }+\delta u^{\mu }, \\
n^{\mu }=\delta n^{\mu },\left. {}\right. \pi ^{\mu \nu }=\delta \pi ^{\mu
\nu }.
\end{gather}%
As already stated, in this analysis we neglect any effects from bulk viscous pressure, i.e., $\delta \Pi =0$. 

Since the fluid 4-velocity is normalized, i.e., $u_{\mu }u^{\mu }=1$, it is straightforward to demonstrate that the
perturbations of the fluid velocity satisfy 
\begin{equation}
u_{0}^{\mu }\delta u_{\mu }=\mathcal{O}\left( 2\right) ,
\end{equation}%
where $\mathcal{O}\left( 2\right) $ denote all possible terms that are second-order or higher in perturbations of the fluid-dynamical fields. That is, up to first-order in perturbations, the fluctuations of the fluid 4-velocity are orthogonal to the background 4-velocity. Similarly, due to the orthogonality relations satisfied by the dissipative currents, Eq.~(\ref{ortho}), one can show that all perturbations of the dissipative currents are also orthogonal to the background 4-velocity, 
\begin{eqnarray}
u_{0}^{\mu }\delta \pi _{\mu \nu } &=&\mathcal{O}\left( 2\right) , \\
u_{0}^{\mu }\delta n_{\mu } &=&\mathcal{O}\left( 2\right) .
\end{eqnarray}%
Such \textit{linear} orthogonality relations satisfied with the background 4-velocity, motivate us to also introduce projection operators that are constructed from the
background 4-velocity,%
\begin{eqnarray}
\Delta _{0}^{\mu \nu } &\equiv &g^{\mu \nu }-u_{0}^{\mu }u_{0}^{\nu }, \\
\Delta _{0}^{\mu \nu \alpha \beta } &\equiv &\frac{1}{2}\left( \Delta
_{0}^{\mu \alpha }\Delta _{0}^{\nu \beta }+\Delta _{0}^{\mu \beta }\Delta
_{0}^{\nu \alpha }\right) -\frac{1}{3}\Delta _{0}^{\mu \nu }\Delta
_{0}^{\alpha \beta },
\end{eqnarray}%
and define the following projected derivatives%
\begin{equation}
D_{0}\equiv u_{0}^{\mu }\partial _{\mu },\left. {}\right. \nabla _{0}^{\mu
}\equiv \Delta _{0}^{\mu \nu }\partial _{\mu }.
\end{equation}

Retaining only contributions which are linear in the perturbations, the fluid-dynamical equations of motion reduce to 
\begin{align}
D_{0}\left( \frac{\delta \varepsilon }{w_{0}}\right) +\nabla _{0}^{\mu
}\delta u_{\mu }& =\mathcal{O}\left( 2\right) , \\
D_{0}\delta u^{\mu }-\nabla _{0}^{\mu }\left( \frac{\delta P}{w_{0}}\right)
+\nabla _{0}^{\nu }\delta \chi _{\nu }^{\mu }& =\mathcal{O}\left( 2\right) ,
\\
D_{0}\left( \frac{\delta n_{B}}{n_{0}}\right) +\nabla _{0}^{\mu }\delta \xi
_{\mu }& =\mathcal{O}\left( 2\right) ,
\end{align}%
where $w_{0}=\varepsilon _{0}+P_{0}$ is the enthalpy and $n_{0}\equiv
(\varepsilon _{0}+P_{0})/4T$ is the particle \textit{number} density at vanishing chemical potential. We further defined the dimensionless fields associated with the hydrodynamic fluctuations, $\delta \chi ^{\mu \nu
}\equiv \delta \pi ^{\mu \nu }/\left( \varepsilon _{0}+P_{0}\right) $ and $%
\delta \xi ^{\mu }\equiv \delta n^{\mu }/n_0$. In this notation, the linearized equations of motion for the dissipative currents become 
\begin{eqnarray}
\tau _{n}D_{0}\delta \xi ^{\mu }+\delta \xi ^{\mu } &=&\frac{\bar{n}_{B}}{n_0}\tau _{\kappa }\nabla
_{0}^{\mu }\delta \alpha _{B}+\mathcal{L}_{n\pi }\nabla _{0}^{\nu }\delta
\chi _{\nu }^{\mu },  \label{linear01} \\
\tau _{\pi }D_{0}\delta \chi ^{\mu \nu }+\delta \chi ^{\mu \nu } &=&2\tau
_{\eta }\Delta _{0}^{\mu \nu \alpha \beta }\partial _{\alpha }\delta
u_{\beta }+\mathcal{L}_{\pi n}\Delta _{0}^{\mu \nu \alpha \beta }\partial
_{\alpha }\delta \xi _{\beta }.  \label{linear02}
\end{eqnarray}%
All transport coefficients in the above equations are functions only of the temperature, since we assumed the chemical potential of the \textit{unperturbed system} is zero. We also introduced several timescales associated with the transport coefficients
that appear in the fluid-dynamical equations,
\begin{equation}
\tau _{\eta }\equiv \frac{\eta }{\varepsilon _{0}+P_{0}},\tau _{\kappa
}\equiv \frac{\kappa _{n}}{\bar{n}_{B}},\left. {}\right. \mathcal{L}_{\pi n}\equiv 
\frac{\ell _{\pi n}}{4T},\left. {}\right. \mathcal{L}_{n\pi }\equiv 4T\ell
_{n\pi },
\end{equation}
with $\bar{n}_{B}$ being the baryon number density, not to be confused with the net-baryon number density. The first two scales appear already in Navier-Stokes theory, while the remaining two are new scales related to the coupling terms investigated in this work. Naturally, there are also intrinsic timescales in Israel-Stewart
theory -- the relaxation times -- which are required to render the theory causal and stable. It would be interesting to see whether the appearance of such new timescales can affect the stability of the linearized theory
(investigating the causality of the full nonlinear theory is a complex task, with recent progress on this topic being developed in Refs.~\cite{jn1, jn2}, but without the effects of net-charge diffusion).

\subsection*{Linearized equations of motion in Fourier space}

It is practical to express these equations in Fourier space. We adopt the following convention for the Fourier transformation 
\begin{eqnarray}
\tilde{M}(k^{\mu }) &=&\int d^{4}x\hspace{0.1cm}\exp \left( -ix_{\mu }k^{\mu
}\right) M(x^{\mu }), \\
M(x^{\mu }) &=&\int \frac{d^{4}k}{(2\pi )^{4}}\hspace{0.1cm}\exp \left(
ix_{\mu }k^{\mu }\right) M(k^{\mu }),
\end{eqnarray}%
where $k^{\mu }=(\omega ,\mathbf{k})$, with $\omega $ being the frequency
and $\mathbf{k}$ the wave vector. It is convenient to introduce the covariant variables, 
\begin{eqnarray}
\Omega &\equiv &u_{0}^{\mu }k_{\mu }, \\
\kappa ^{\mu } &\equiv &\Delta _{0}^{\mu \nu }k_{\nu },
\end{eqnarray}%
which correspond to the frequency and wave vector in the local rest frame of the unperturbed system. We further introduce a covariant wave number,
\begin{equation}
\kappa\equiv \sqrt{-\kappa_{\mu}\kappa^{\mu}}.
\end{equation}
Using this notation, the conservation laws are written as%
\begin{align}
\Omega \frac{\delta \tilde{\varepsilon}}{w_{0}}+\kappa ^{\mu }\delta \tilde{u%
}_{\mu }& =0,  \label{bla1} \\
\Omega \delta \tilde{u}^{\mu }-\kappa ^{\mu }\frac{\delta \tilde{P}}{w_{0}}%
+\kappa ^{\nu }\delta \tilde{\chi}_{\nu }^{\mu }& =0,  \label{bla2} \\
\Omega \frac{\delta \tilde{n}_{B}}{n_{0}}+\kappa ^{\mu }\delta \tilde{\xi}%
_{\mu }& =0,  \label{bla3}
\end{align}%
while the equations for the dissipative currents become 
\begin{eqnarray}
(i\tau _{n}\Omega +1)\delta \tilde{\xi}^{\mu } &=&i\frac{\bar{n}_{B}}{n_0}\tau _{\kappa }\kappa
^{\mu }\delta \tilde{\alpha}_{B}+i\mathcal{L}_{n\pi }\kappa _{\nu }\delta 
\tilde{\chi}^{\mu \nu },  \label{ntilde} \\
(i\tau _{\pi }\Omega +1)\delta \tilde{\chi}^{\mu \nu } &=&2i\tau _{\eta }%
\left[ \kappa ^{(\mu }\delta \tilde{u}^{\nu )}-\frac{1}{3}\Delta _{0}^{\mu
\nu }\kappa _{\lambda }\delta \tilde{u}^{\lambda }\right] +i\mathcal{L}_{\pi
n}\left[ \kappa ^{(\mu }\delta \tilde{\xi}^{\nu )}-\frac{1}{3}\Delta ^{\mu
\nu }\kappa _{\lambda }\delta \tilde{\xi}^{\lambda }\right] ,  \label{ptilde}
\end{eqnarray}%
where the parentheses in the indices denote the symmetrized tensor $A^{(\mu
\nu )}\equiv \left( A^{\mu \nu }+A^{\nu \mu }\right) /2$.

For the sake of convenience, we further decompose the above equations of
motion in terms of their components that are parallel to $\kappa ^{\mu }$ (\textit{longitudinal degrees of freedom}) and those that are orthogonal to it (\textit{transverse degrees of freedom}). In the linear regime, the transverse and longitudinal degrees of freedom are no longer coupled and can be solved independently. For instance, the decomposition of an arbitrary 4-vector orthogonal to $u^\mu_0$, $A^{\mu }$, can be implemented as
\begin{equation}
A^{\mu }=A_{\Vert }\frac{\kappa ^{\mu }}{\kappa }+A_{\bot }^{\mu },
\end{equation}%
where we define $A_{\Vert }\equiv -\kappa _{\mu }A^{\mu }/\kappa $ and $%
A_{\bot }^{\mu }\equiv \Delta _{\kappa }^{\mu \nu }A_{\nu }$. Here, 
\begin{equation}
\Delta
_{\kappa }^{\mu \nu }\equiv g^{\mu \nu }+\frac{\kappa ^{\mu }\kappa ^{\nu }}{\kappa^{2}}-u^\mu_0 u^\nu_0,
\end{equation}
is the orthogonal projector onto the 2-space orthogonal to $\kappa
^{\mu }$ and $u^\mu_0$. A symmetric traceless rank two tensor orthogonal to $u^\mu_0$ can be decomposed in a similar manner%
\begin{equation}
A^{\mu \nu }=A_{\Vert }\frac{\kappa ^{\mu }\kappa ^{\nu }}{\kappa ^{2}}+%
\frac{1}{2}A_{\Vert }\Delta _{\kappa }^{\mu \nu }+A_{\bot }^{\mu }\frac{%
\kappa ^{\nu }}{\kappa }+A_{\bot }^{\nu }\frac{\kappa ^{\mu }}{\kappa }%
+A_{\bot }^{\mu \nu },
\end{equation}%
where we define the projections $A_{\Vert }\equiv \kappa _{\mu }\kappa _{\nu
}A^{\mu \nu }/\kappa ^{2}$, $A_{\bot }^{\mu }\equiv -\kappa ^{\lambda
}\Delta _{\kappa }^{\mu \nu }A_{\lambda \nu }/\kappa $, and $A_{\bot }^{\mu
\nu }\equiv \Delta _{\kappa }^{\mu \nu \alpha \beta }A_{\alpha \beta }$,
with 
\begin{equation}
\Delta _{\kappa }^{\mu \nu \alpha \beta }\equiv \frac{1}{2}\left( \Delta
_{\kappa }^{\mu \alpha }\Delta _{\kappa }^{\nu \beta }+\Delta _{\kappa
}^{\mu \beta }\Delta _{\kappa }^{\nu \alpha } -\Delta
_{\kappa }^{\mu \nu }\Delta _{\kappa }^{\alpha \beta }\right).
\end{equation}

The equations of motion satisfied by the longitudinal modes are obtained by
contracting Eqs. (\ref{bla2}) and (\ref{ntilde}) with the tensor $\kappa
^{\mu }/\kappa $ and by contracting Eq. (\ref{ptilde}) with the tensor $%
\kappa ^{\mu }\kappa ^{\nu }/\kappa ^{2}$. Note that Eqs. (\ref{bla1}) and (%
\ref{bla3}) are already in terms of the longitudinal components of the
fluctuations. This leads to the equations%
\begin{align}
\Omega \frac{\delta \tilde{\varepsilon}}{w_{0}}-\kappa \delta \tilde{u}%
_{\Vert }& =0, \\
\Omega \delta \tilde{u}_{\Vert }-\kappa \frac{\delta \tilde{\varepsilon}}{%
3w_{0}}-\kappa \delta \tilde{\chi}_{\Vert }& =0, \\
\Omega \frac{\delta \tilde{n}_{B}}{n_{0}}-\kappa \delta \tilde{\xi}_{\Vert
}& =0, \\
\left( i\hat{\tau}_{n}\hat{\Omega}+1\right) \delta \tilde{\xi}_{\Vert }+i%
\mathcal{\hat{L}}_{n\pi }\hat{\kappa}\delta \tilde{\chi}_{\Vert }& =i\hat{%
\tau}_{\kappa }\hat{\kappa}\frac{\delta \tilde{n}_{B}}{n_{0}}, \\
\left( i\hat{\tau}_{\pi }\hat{\Omega}+1\right) \delta \tilde{\chi}_{\Vert }-%
\frac{2}{3}i\mathcal{\hat{L}}_{\pi n}\hat{\kappa}\delta \tilde{\xi}_{\Vert
}& =\frac{4}{3}i\hat{\kappa}\delta \tilde{u}_{\Vert },
\end{align}%
Above, we expressed all dimensionful quantities
in terms of the viscous timescale, $\tau _{\eta }$. That is, we defined
\begin{eqnarray}
\hat{\Omega} &\equiv &\tau _{\eta }\Omega ,\left. {}\right. \hat{\kappa}%
\equiv \tau _{\eta }\kappa ,\left. {}\right. \hat{\tau}_{n,\pi ,\kappa
}\equiv \frac{\tau _{n,\pi ,\kappa }}{\tau _{\eta }}, \\
\mathcal{\hat{L}}_{n\pi } &\equiv &\frac{\mathcal{L}_{n\pi }}{\tau _{\eta }}%
,\left. {}\right. \mathcal{\hat{L}}_{\pi n}\equiv \frac{\mathcal{L}_{\pi n}}{%
\tau _{\eta }}.
\end{eqnarray}
In deriving the above equations, we have already made assumptions regarding the equation of state. We assume an equation of state of a gas composed solely of a massless particle and its corresponding antiparticle. In this case, the perturbation of pressure and chemical potential can be expressed as
\begin{eqnarray}
\delta \tilde{P} &=&\frac{1}{3}\delta \tilde{\varepsilon}, \\
\delta \tilde{\alpha}_{B} &=&\frac{\delta \tilde{n}_{B}}{\bar{n}_{B}}.
\end{eqnarray}

The transverse equations are obtained projecting Eqs.~(\ref{bla2}) and (\ref%
{ntilde}) with $\Delta_{\kappa }^{\mu \lambda }$ and Eq.~(\ref{ptilde}) with $\Delta _{\kappa }^{\mu \lambda }\kappa^\nu$. Then, we
obtain the following set of equations%
\begin{eqnarray}
\hat{\Omega}\delta \tilde{u}_{\bot }^{\lambda }-\hat{\kappa}\delta \tilde{\chi}%
_{\bot }^{\lambda } &=&0, \\
(i\hat{\tau}_{n}\hat{\Omega}+1)\delta \tilde{\xi}_{\bot }^{\lambda }+i\mathcal{%
\hat{L}}_{n\pi }\hat{\kappa}\delta \tilde{\chi}_{\bot }^{\lambda } &=&0, \\
(i\hat{\tau}_{\pi }\hat{\Omega}+1)\delta \tilde{\chi}_{\bot }^{\lambda }-i\hat{%
\kappa}\delta \tilde{u}_{\bot }^{\lambda }-i\frac{\mathcal{\hat{L}}_{\pi n}}{2}%
\hat{\kappa}\delta \tilde{\xi}_{\bot }^{\lambda } &=&0.
\end{eqnarray}
The equation for the fully transverse component of the shear-stress tensor, $\delta\tilde\chi^{\mu\nu}_\bot$, decouples from energy density and velocity fluctuations and will not be considered in this analysis. 

When required, we use the transport coefficients calculated in Ref.~\cite{dnmr} for an ultrarelativistic gas of hard spheres. In this case, one finds for the rescaled transport coefficients the following values: $\hat{\tau}_{\pi }=5$, $\hat{\tau}_{n}=27/4$, $\hat{\tau}_{\kappa}=9/16$. The coefficients related to the coupling terms, $\mathcal{L}_{n\pi}$ and $\mathcal{L}_{\pi n}$, will not be fixed to a particular value. Nevertheless, we shall assume in this work, unless stated otherwise, that
\begin{equation}
\mathcal{L}_{n\pi}\mathcal{L}_{\pi n}<0.
\end{equation}
This assumption is supported by kinetic theory calculations \cite{is2, dnmr, 14moment}. Furthermore, this constraint is obtained in the phenomenological derivation of Israel-Stewart theory from the second law of thermodynamics \cite{is2, muronga}.

\section{Stability analysis without coupling terms}
\label{stab_wo}

We begin our analysis considering the case where coupling terms between the shear-stress tensor and diffusion 4-current are not present. This analysis
was performed before, without the presence of diffusion fluctuations in Ref.~\cite{rischke}. Preliminary calculations on the effect of diffusion fluctuations have been presented in Ref.~\cite{pushi}.

\subsection{Transverse modes}

We start our discussion with the transverse degrees of freedom. In this case, the equations of motion, taking $\mathcal{\hat{L}}_{n\pi }=\mathcal{\hat{L}}_{\pi n}=0$, can be expressed in the following form 
\begin{equation}
\left( 
\begin{array}{ccc}
\hat{\Omega} & -\hat{\kappa} & 0 \\ 
0 & 0 & i\hat{\Omega}\hat{\tau}_{n}+1 \\ 
-i\hat{\kappa} & i\hat{\tau}_{\pi }\hat{\Omega}+1 & 0%
\end{array}%
\right) \left( 
\begin{array}{c}
\delta \tilde{u}_{\bot }^{\mu } \\ 
\delta \tilde{\chi}_{\bot }^{\mu } \\ 
\delta \tilde{\xi}_{\bot }^{\mu }%
\end{array}%
\right) =0.
\end{equation}%
The nontrivial solutions are obtained when the determinant vanishes,
leading to the dispersion relation
\begin{equation}
\left[ \left( 1+i\hat{\tau}_{\pi }\hat{\Omega}\right) \hat{\Omega}-i\hat{%
\kappa}^{2}\right] \left( 1+i\hat{\tau}_{n}\hat{\Omega}\right) =0.  \notag
\end{equation}%
Note that, in the absence of coupling terms, the transverse dispersion
relations related to the shear-stress tensor and net-baryon diffusion current
decouple, and can be expressed as
\begin{eqnarray}
\left( 1+i\hat{\tau}_{\pi }\hat{\Omega}\right) \hat{\Omega}-i\hat{\kappa}%
^{2} &=&0,  \label{DispShear} \\
1+i\hat{\tau}_{n}\hat{\Omega} &=&0.  \label{DispDiff}
\end{eqnarray}
Such decoupling of the modes would not necessarily occur for other choices of the equation of state. Also, this will not occur if we consider fluctuations around an equilibrium state with a finite net-baryon number density.

Initially, we shall consider the case where the unperturbed fluid is at
rest, i.e., $u_{0}^{\mu }=\left(1,0,0,0\right) $, leading to $\hat{\Omega}=%
\hat{\omega}$ and $\hat{\kappa}=\hat{k}$. In this case, the solutions can be
expressed analytically as 
\begin{equation}
\hat{\omega}_{T}^{\mathrm{diff}}=\frac{i}{\hat{\tau}_{n}},\left. {}\right. 
\hat{\omega}_{T,\pm }^{\mathrm{shear}}=i\frac{1\pm \sqrt{1-4\hat{\tau}_{\pi }%
\hat{k}^{2}}}{2\hat{\tau}_{\pi }}.
\end{equation}%
Clearly, the modes $\hat{\omega}_{T}^{\mathrm{diff}}$ and $\hat{\omega}%
_{T,+}^{\mathrm{shear}}$ are nonhydrodynamic, i.e., they do not vanish when the wave number is taken to zero, while the mode $\hat{\omega}_{T,-}^{\mathrm{%
shear}}$ is hydrodynamic. Also, it is straightforward to see that, when
taking the Navier-Stokes limit, i.e., when the relaxation times are sent to
zero, $\hat{\tau}_{\pi }\rightarrow 0$, $\hat{\tau}_{n}\rightarrow 0$, the
nonhydrodynamic modes go to infinity, $\lim_{\hat{\tau}_{n}\rightarrow 0}%
\hat{\omega}_{T}^{\mathrm{diff}}\sim i/\hat{\tau}_{n}$ and $\lim_{\hat{\tau}%
_{\pi }\rightarrow 0}\hat{\omega}_{T,+}^{\mathrm{shear}}\sim i/\hat{\tau}%
_{\pi }$, while the hydrodynamic mode goes to the
usual Navier-Stokes solution, $\lim_{\hat{\tau}_{\pi }\rightarrow 0}\hat{%
\omega}_{T,-}^{\mathrm{shear}}=i\hat{k}^{2}$. We also see that the shear
modes become propagating; i.e., they have a real component, for wave numbers
that are sufficiently large, $\hat{k}>1/\left( 2\sqrt{\hat{\tau}_{\pi }}%
\right) $. All these results for the shear modes were previously obtained in
Ref.~\cite{rischke}.

Since they are the only ones that carry any dependence on the wave number $\hat{k}$, it is useful to look at the modes $\hat{\omega}^\mathrm{shear}_{T,\pm}$ in the small, $\hat{k}\ll 1$, and
large, $\hat{k}\gg 1$, wave number limits. In the first case, we have
\begin{eqnarray}
\hat{\omega}_{T,+}^{\mathrm{shear}} &=&\frac{i}{\hat{\tau}_{\pi }}-i\hat{k}%
^{2}+\mathcal{O}\left( \hat{k}^{4}\right) , \\
\hat{\omega}_{T,-}^{\mathrm{shear}} &=&i\hat{k}^{2}+i\hat{\tau}_{\pi }\hat{k}%
^{4}+\mathcal{O}\left( \hat{k}^{6}\right) ,
\end{eqnarray}%
while, in the second case, we obtain%
\begin{equation}
\hat{\omega}_{T,\pm }^{\mathrm{shear}}=\frac{i}{2\hat{\tau}_{\pi }}\pm
\left( \frac{\hat{k}}{\sqrt{\hat{\tau}_{\pi }}}-\frac{1}{8\hat{\tau}_{\pi
}^{3/2}\hat{k}}\right) +\mathcal{O}\left( \frac{1}{\hat{k}^{4}}\right).
\end{equation}%
We can see that the modes obtained above are stable as long as the
relaxation times are positive%
\begin{equation}
\hat{\tau}_{n}\geq 0,\hat{\tau}_{\pi }\geq 0.
\end{equation}%
Furthermore, since the shear modes become propagating when $\hat{k}\gg 1$, causality imposes the following constraint to the asymptotic group velocity \cite{rischke, jackson}
\begin{equation}
\lim_{\hat{k}\rightarrow \infty }\left\vert\frac{\partial\mathrm{Re}(\hat{\omega})}{\partial\hat{k}}\right\vert \leq1\Longrightarrow \hat{\tau}_{\pi }\geq 1.\label{causal01}
\end{equation}

\begin{figure}[ht]
\begin{center}
\includegraphics[scale=0.52]{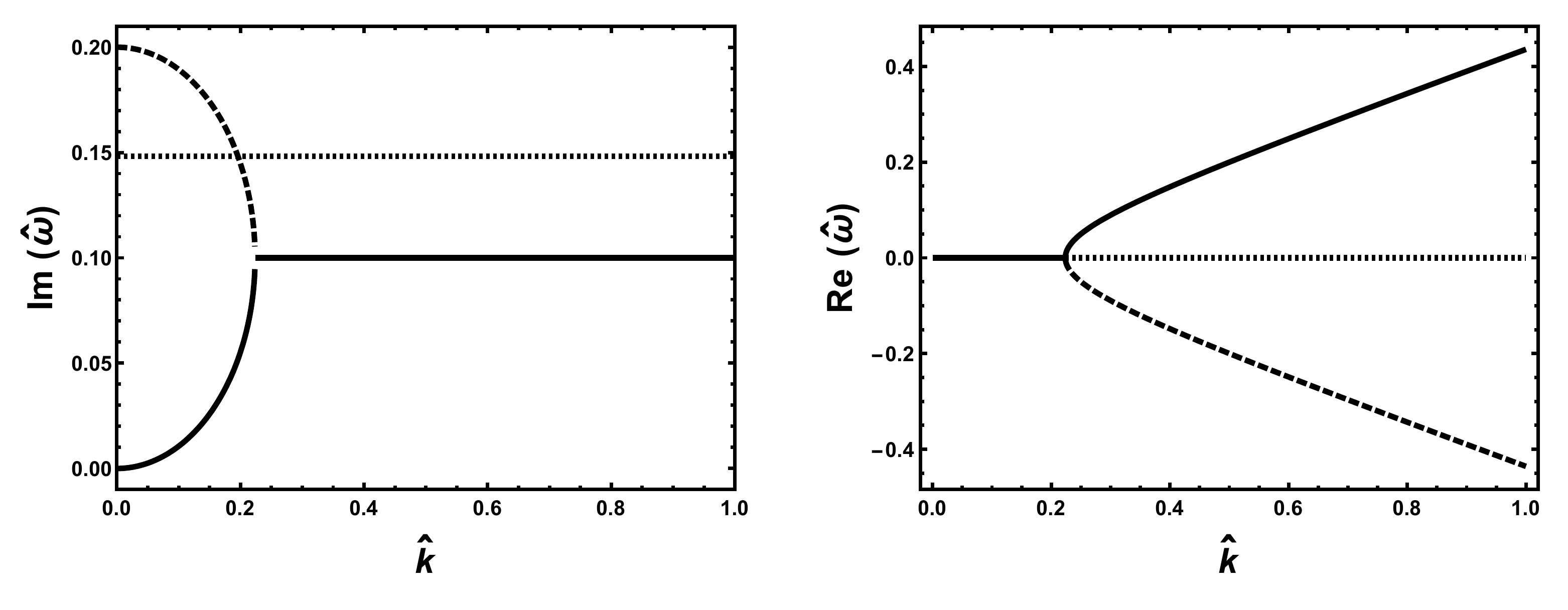}
\caption{The imaginary and real parts of the transverse modes in the absence of coupling terms, i.e., $\hat{\mathcal{L}}_{n\pi}\hat{\mathcal{L}}_{n\pi}=0$.}
\label{fig1_trans}
\end{center}
\end{figure}

The modes $\hat{\omega}_{T,\pm }^{\mathrm{shear}}$ and $\hat{\omega}_{T}^{\mathrm{diff}}$ are plotted in Fig.~\ref{fig1_trans}, considering the relaxation times
calculated from the Boltzmann equation, using the 14-moment approximation in
the ultrarelativistic limit, i.e., $\hat{\tau}_{\pi }=5$ and $\hat{\tau}%
_{n}=27/4$ \cite{dnmr}. The results do not change qualitatively if different values of $%
\hat{\tau}_{\pi }$ and $\hat{\tau}_{n}$ are employed. The transverse modes
related to the shear-stress tensor are usually referred to as shear modes.
These modes have been calculated before in Ref.~\cite{rischke} and are identical to
the results presented here, since the inclusion of net-baryon diffusion does
not affect them. 

We now analyze the case in which the unperturbed fluid is moving. Such problem involves intrinsically relativistic phenomena, since velocities that are close to the velocity of light are possible even in such perturbative scenario. We define the Lorentz factor, $\gamma =1/\sqrt{1-V^{2}}$, and, for the sake of simplicity, assume that the 3-velocity and the wave vector are taken in the same direction, e.g., the $x$ axis. This case corresponds to $u_{0}^{\mu }=\gamma (1,V,0,0)$
and $k^{\mu }=(\omega ,k,0,0)$, leading to 
\begin{eqnarray}
\hat{\Omega} &=&\gamma (\hat{\omega}-V\hat{k}),  \label{omega} \\
\hat{\kappa}^2 &=&\gamma^2(\hat{\omega} V-\hat{k})^2.  \label{kappa}
\end{eqnarray}

In this case, the dispersion relations given by Eqs. (\ref{DispShear}) and (\ref{DispDiff}) become 
\begin{eqnarray}
i\hat{\tau}_{\pi }(\gamma \hat{\omega}-\gamma V\hat{k})^{2}+(\gamma \hat{%
\omega}-\gamma V\hat{k})-i(\gamma \hat{\omega}V-\gamma \hat{k})^{2} &=&0, \\
1+i\gamma \hat{\tau}_{n}(\hat{\omega}-V\hat{k}) &=&0.
\end{eqnarray}%
The solution for the diffusion transverse mode is, once more, the simplest
one,%
\begin{equation}
\hat{\omega}_{T}^{\mathrm{diff}}=V\hat{k}+\frac{i}{\gamma \hat{\tau}_{n}}
\end{equation}%
given by a propagating part, with velocity identical to that of the
unperturbed fluid, and a nonpropagating part  that relaxes to equilibrium within a Lorentz dilated relaxation time, $\gamma \hat{\tau}_{n}$ (this corresponds to a nonhydrodynamic mode). This mode is always stable as long as the diffusion relaxation time is positive. Furthermore, as expected, causality is satisfied as long as the background fluid has a velocity smaller than the speed of light. The dispersion relation satisfied by the remaining transverse modes is,

\begin{equation}
i\left( \hat{\tau}_{\pi }-V^{2}\right) \left( \gamma \hat{\omega}\right)
^{2}+\left[ 1-2i\left( \hat{\tau}_{\pi }-1\right) \gamma V\hat{k}\right]
\gamma \hat{\omega}-\gamma V\hat{k}+i( \hat{\tau}_{\pi }V^{2}-1)
( \gamma \hat{k}) ^{2}=0.\label{disper01}
\end{equation}%
In order to understand the stability of such modes, we first look at their behavior when $\hat{k}=0$, i.e., in the homogeneous limit. In this limiting case, the hydrodynamic mode will naturally disappear,
while the nonhydrodynamic mode achieves the following value 
\begin{equation}
\hat{\omega}_{T,+}^{\mathrm{shear}}(\hat{k}=0) =\frac{i}{\gamma
\left( \hat{\tau}_{\pi }-V^{2}\right) }.
\end{equation}
We note that this mode does not vanish when the shear relaxation time is taken to zero. This implies that such nonhydrodynamic mode also appears in Navier-Stokes theory. 

The nonhydrodynamic mode should be stable for all possible values of the background velocity $V$. In order to guarantee this at least for $k=0$, the shear relaxation time must satisfy the following condition
\begin{equation}
\hat{\tau}_{\pi }\geq 1 \Longrightarrow \tau_{\pi }\geq \tau_{\eta }. 
\end{equation}%
This constraint is identical to the causality condition obtained for the same modes when the unperturbed fluid was at rest; see Eq.~(\ref{causal01}). This linear stability condition for the shear modes was previously obtained by Pu \textit{et al.} in Ref.~\cite{rischke}, and can be shown to be equivalent to one of the constraints for linear stability obtained by Olson in Ref.~\cite{olson} [stability condition (65) applied to Eq.~(52) of the aforementioned paper in the limit of vanishing density].
\begin{figure}[ht]
\begin{center}
\includegraphics[scale=0.35]{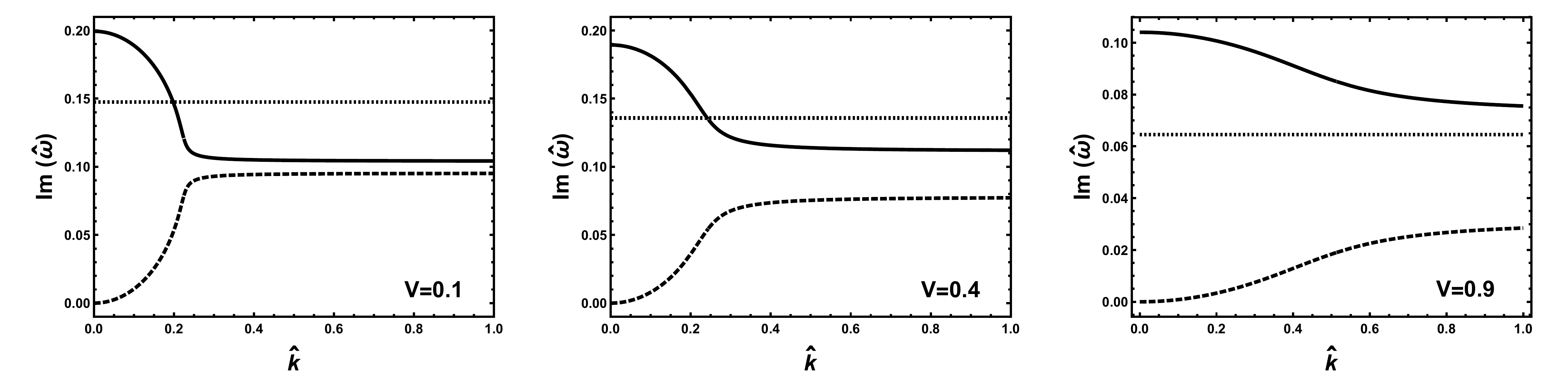} \\
\includegraphics[scale=0.35]{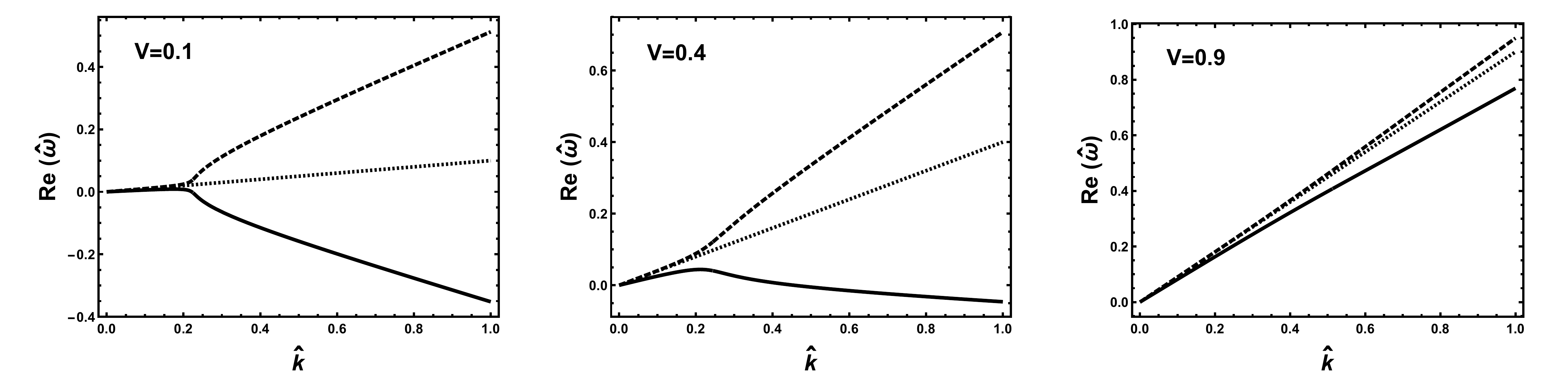}
\caption{The imaginary and real parts of the transverse modes for perturbations around a moving fluid, considering $V=0.1$, $V=0.4$, and $V=0.9$, in the absence of coupling terms, i.e., $\hat{\mathcal{L}}_{n\pi}\hat{\mathcal{L}}_{n\pi}=0$.}
\label{fig2_trans}
\end{center}
\end{figure}

\begin{figure}[ht]
\begin{center}
\includegraphics[scale=0.45]{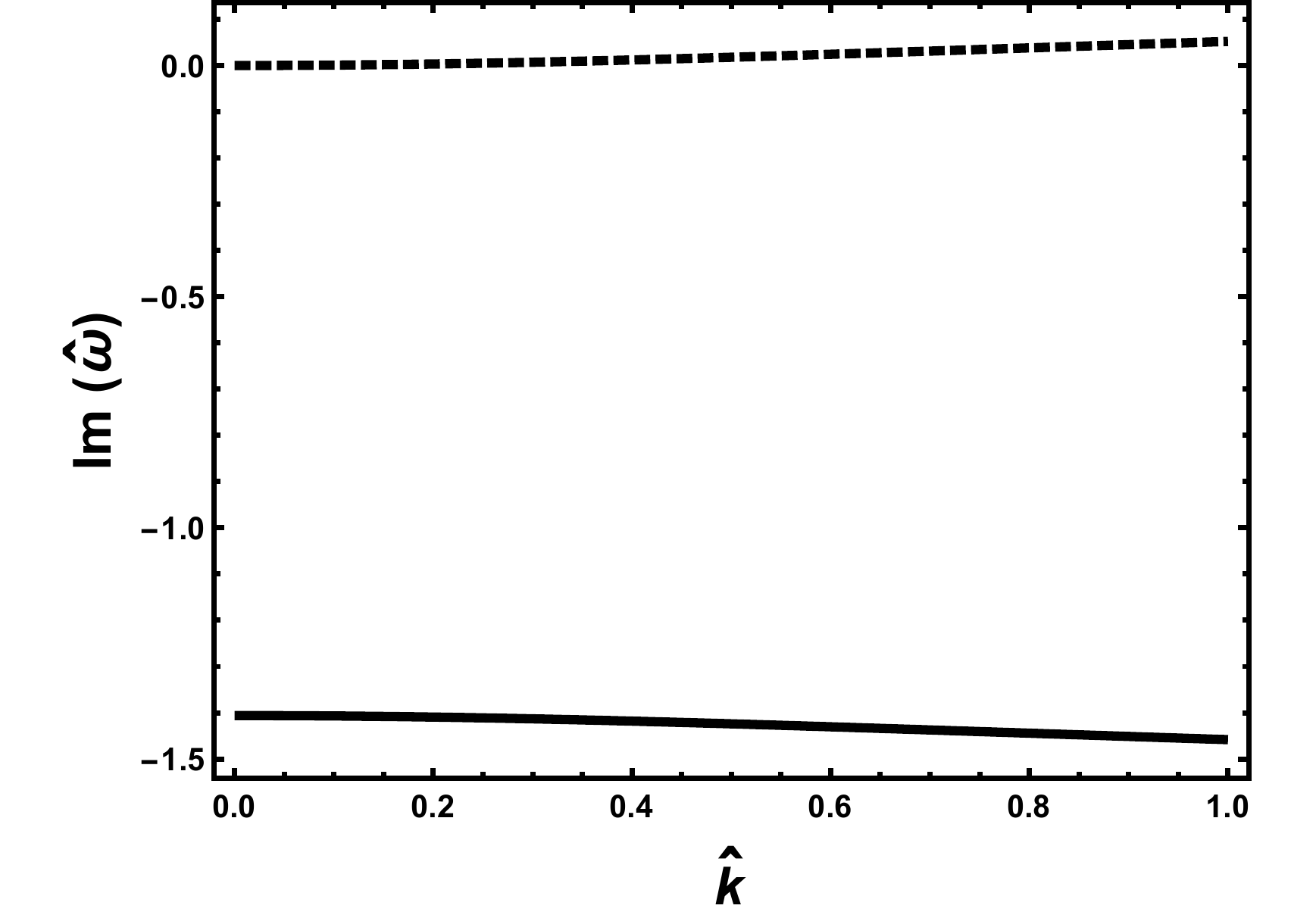}
\caption{The imaginary part of the unstable shear mode for $\hat{\tau}_{\pi}=0.5$ for a background velocity of $V=0.9$, in the absence of coupling terms, i.e., $\hat{\mathcal{L}}_{n\pi}\hat{\mathcal{L}}_{n\pi}=0$.}
\label{fig4uns}
\end{center}
\end{figure}

In Fig.~\ref{fig2_trans}, we plot the solutions of Eq.~(\ref{disper01}) for $V=0.1$, $V=0.4$, and $V=0.9$, using the same values of relaxation times as before: $\hat{\tau}_{\pi }=5$ and $\hat{%
\tau}_{n}=27/4$. In this case, we see that the modes are indeed stable for all values of $\hat k$. An example of an unstable fluid configuration is shown in Fig.~\ref{fig4uns}, by taking $\hat{\tau}_{\pi}=0.5$ and $\hat{\tau}_n=27/4$, values that do not satisfy the causality and stability conditions derived in this section for the transverse modes. 

\subsection{Longitudinal modes}

We now discuss the longitudinal projections of the fluid-dynamical equations. The equations of motion for the longitudinal degrees of freedom, taking $\hat{\mathcal{L}}_{n\pi}=\hat{\mathcal{L}}_{n\pi}=0$, can be expressed in the following matrix form
\begin{equation}
\left( 
\begin{array}{ccccc}
\Omega & 0 & 0 & -\kappa & 0 \\ 
0 & \Omega & -\kappa & 0 & 0 \\ 
0 & -\frac{\kappa}{3} & \Omega & 0 & -\kappa \\ 
-i\hat{\tau}_{\kappa }\hat{\kappa} & 0 & 0 & i\hat{\tau}_{n}\hat{\Omega}+1 & 
0 \\ 
0 & 0 & -\frac{4}{3}i\hat{\kappa} & 0 & i\hat{\tau}_{\pi }\hat{\Omega}+1%
\end{array}%
\right) \left( 
\begin{array}{c}
\delta \tilde{n}_{B}/n_{0} \\ 
\delta \tilde{\varepsilon}/w_{0} \\ 
\delta \tilde{u}_{\Vert } \\ 
\delta \tilde{\xi}_{\Vert } \\ 
\delta \tilde{\chi}_{\Vert }%
\end{array}%
\right) =0.
\end{equation}%
Nontrivial solutions for these equations are obtained when the determinant is zero, leading to the dispersion relation
\begin{equation}
\left[ \left( \hat{\Omega}^{2}-\frac{1}{3}\hat{\kappa}^{2}\right) (i\hat{\tau%
}_{\pi }\hat{\Omega}+1)-\frac{4}{3}i\hat{\kappa}^{2}\hat{\Omega}\right] %
\left[ \hat{\Omega}(i\hat{\tau}_{n}\hat{\Omega}+1)-i\hat{\tau}_{\kappa }\hat{%
\kappa}^{2}\right] =0.\label{disper02}
\end{equation}%
Once more, the dispersion relation related to net-baryon current perturbations decouples from those related to energy-momentum tensor fluctuations and can be solved independently,
\begin{eqnarray}
\hat{\Omega}(i\hat{\tau}_{n}\hat{\Omega}+1)-i\hat{\tau}_{\kappa }\hat{\kappa}%
^{2} &=&0,  \label{DispDiffL} \\
\left( \hat{\Omega}^{2}-\frac{1}{3}\hat{\kappa}^{2}\right) (i\hat{\tau}_{\pi
}\hat{\Omega}+1)-\frac{4}{3}i\hat{\kappa}^{2}\hat{\Omega} &=&0.
\label{DispShearL}
\end{eqnarray}

As before, we first consider the case where the unperturbed fluid is at rest, i.e., $u_{0}^{\mu }=\left( 1,0,0,0\right) $, leading to $\hat{\Omega}=\hat{\omega}$ and $\hat{\kappa}=\hat{k}$. In this case, the solution for the modes related to the net-baryon diffusion current fluctuations can be cast in a simple form
\begin{equation}
\omega _{L,\pm }^{\mathrm{B}}=i\frac{1\pm \sqrt{1-4\hat{\tau}_{n}\hat{\tau%
}_{\kappa }\hat{k}^{2}}}{2\hat{\tau}_{n}}.
\end{equation}%
The mode $\omega _{L,-}^{\mathrm{B}}$ is clearly hydrodynamic, as can be seen from the small-wave-number limit
\begin{equation}
\omega _{L,-}^{\mathrm{B}}=i\hat{\tau}_{\kappa }\hat{k}^{2}+i\hat{\tau}%
_{n}\hat{\tau}_{\kappa }^{2}\hat{k}^{4}+\mathcal{O}\left( \hat{k}^{6}\right),
\end{equation}%
with the leading term being the dispersion related usually obtained in Navier-Stoke theory. On the other hand, $\omega _{L,+}^{\mathrm{B}}$ is a nonhydrodynamic mode that simply does not exist in Navier-Stokes theory,%
\begin{equation}
\omega _{L,+}^{\mathrm{B}}=\frac{i}{\hat{\tau}_{n}}-i\hat{\tau}_{\kappa }%
\hat{k}^{2}+\mathcal{O}\left( \hat{k}^{4}\right) .
\end{equation}%
For asymptotically large values of wave number, these modes become%
\begin{equation}
\omega _{L,\pm }^{\mathrm{B}}=\frac{i}{2\hat{\tau}_{n}}\pm \sqrt{\frac{%
\hat{\tau}_{\kappa }}{\hat{\tau}_{n}}}\left( \hat{k}-\frac{1}{8\hat{\tau}_{n}%
\hat{\tau}_{\kappa }\hat{k}}\right) +\mathcal{O}\left( \frac{1}{\hat{k}^{3}}%
\right) .
\end{equation}
The modes $\omega _{L,\pm }^{\mathrm{diff}}$ become propagating when $\hat{k}>1/\left( 2\sqrt{\hat{%
\tau}_{n}\hat{\tau}_{\kappa }}\right) $ and causality dictates that \cite{rischke, jackson}
\begin{equation}
\lim_{k\rightarrow \infty}\left\vert\frac{\partial \mathrm{Re}(\omega)}{\partial k}\right\vert\leq 1\Longrightarrow \hat{\tau}_{n}\geq \hat{\tau}_{\kappa }.\label{causal02}
\end{equation}

\begin{figure}[ht]
\begin{center}
\includegraphics[scale=0.52]{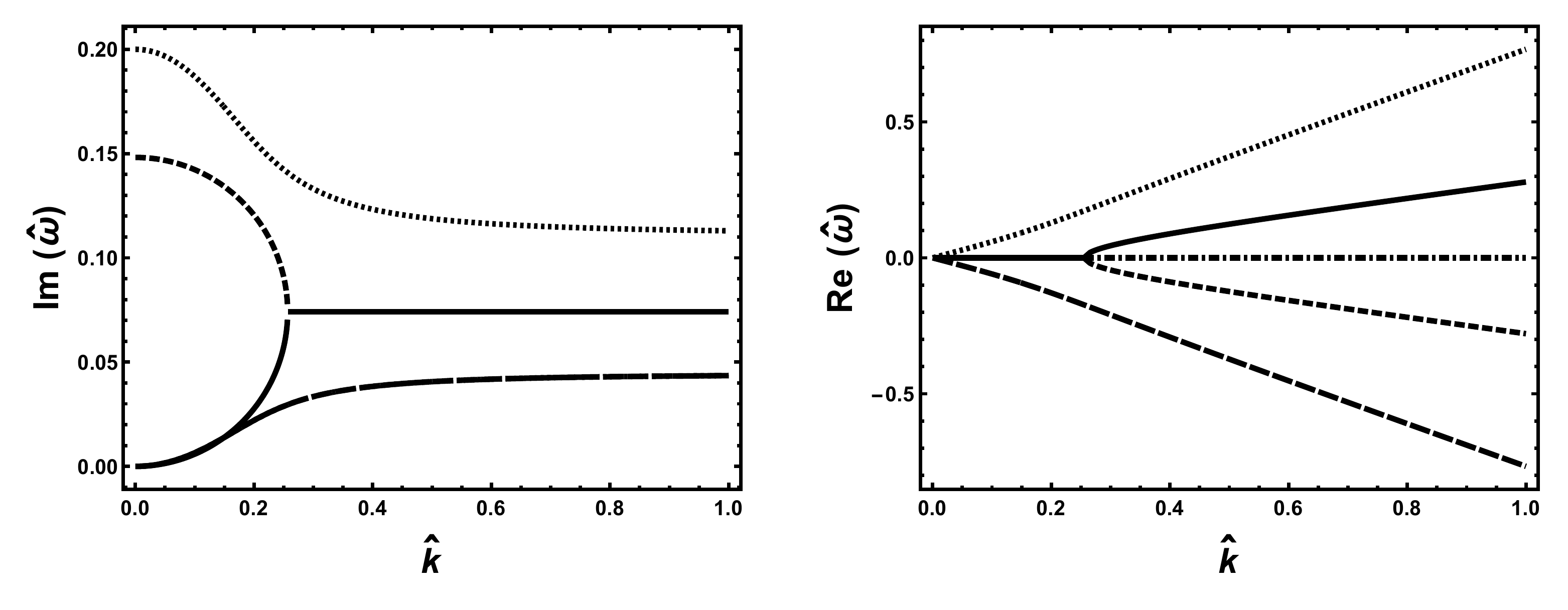}
\caption{Real and imaginary part of the longitudinal modes in the absence of coupling terms, $\hat{\mathcal{L}}_{n\pi}\hat{\mathcal{L}}_{n\pi}=0$, for a static background, $V=0$.}
\label{long1}
\end{center}
\end{figure}

The remaining longitudinal modes (which include the sound modes) are a solution of a cubic equation and, thus, their analytical solution cannot be expressed in a simple form. For the sake of completeness, we plot the imaginary and real parts of the longitudinal modes in Fig.~\ref{long1}, taking $\hat{\tau}_{\pi}=5$, $\hat{\tau}_{n}=27/4$, and $\hat{\tau}_\kappa=9/16$. In the following, we restrict our discussion to the asymptotic form of these modes. We take a look at the behavior of such modes for small values of wave number, 
\begin{eqnarray}
\omega _{\pm }^{\mathrm{sound}} &=&\pm \frac{1}{3}\hat{k}+\frac{2}{3}i\hat{%
k}^{2}+\mathcal{O}\left( \hat{k}^{3}\right) , \\
\omega _{L}^{\mathrm{shear}} &=&\frac{i}{\hat{\tau}_{\pi }}-\frac{4}{3}i%
\hat{k}^{2}+\mathcal{O}\left( \hat{k}^{4}\right) ,
\end{eqnarray}%
while at large values of wave number, the same modes behave as
\begin{eqnarray}
\omega _{\pm }^{\mathrm{sound}} &=&\pm \sqrt{\frac{4+\hat{\tau}_{\pi }}{3%
\hat{\tau}_{\pi }}}\hat{k}+\frac{2i}{\hat{\tau}_{\pi }\left( 4+\hat{\tau}%
_{\pi }\right) }+\mathcal{O}\left( \frac{1}{\hat{k}}\right) , \\
\omega _{L}^{\mathrm{shear}} &=&\frac{i}{4+\hat{\tau}_{\pi }}+\mathcal{O}%
\left( \frac{1}{\hat{k}}\right) .
\end{eqnarray}%
Clearly, the modes $\hat{\omega}^\mathrm{sound}_{\pm}$ are hydrodynamic (describing the propagation of sound waves) and, as expected, they reduce to the modes obtained from Navier-Stokes theory when the wave number is sufficiently small. The other mode is nonhydrodynamic and does not exist in Navier-Stokes theory. The stability of these modes is guaranteed for small or large values of wave number as long as $\hat{\tau}_{\pi }>0$.
Furthermore, causality imposes that the asymptotic group velocity must satisfy \cite{rischke, jackson}
\begin{equation}
\lim_{k\rightarrow \infty}\left\vert\frac{\partial \mathrm{Re}(\omega)}{\partial k}\right\vert\leq1\Longrightarrow\hat{\tau}_{\pi }\geq 2.\label{causal03}
\end{equation}

We now perform the same analysis considering that the unperturbed fluid is moving. As already done in the previous section, we assume that the background 4-velocity and the
perturbations are in the same direction, e.g., the $x$ axis.
Therefore, we have $u_{0}^{\mu}=\gamma(1,V,0,0)$ and $k^{\mu
}=(\omega ,k,0,0)$. In this case, the original dispersion relations, Eqs.~(\ref{DispDiffL}) and (\ref{DispShearL}), become
\begin{eqnarray}
\left( \gamma \hat{\omega}-\gamma V\hat{k}\right) \left[ i\hat{\tau}
_{n}\left( \gamma \hat{\omega}-\gamma V\hat{k}\right) +1\right] -i\hat{\tau}%
_{\kappa }\left( \hat{\omega}\gamma V-\gamma \hat{k}\right) ^{2} &=&0, \\
\left[ \left( \gamma \hat{\omega}-\gamma V\hat{k}\right) ^{2}-\frac{1}{3}%
\left( \hat{\omega}\gamma V-\gamma \hat{k}\right) ^{2}\right] \left[ i\hat{%
\tau}_{\pi }\left( \gamma \hat{\omega}-\gamma V\hat{k}\right) +1\right] -%
\frac{4}{3}i\left( \hat{\omega}\gamma V-\gamma \hat{k}\right) ^{2}\left(
\gamma \hat{\omega}-\gamma V\hat{k}\right) &=&0.
\end{eqnarray}

Since analyzing the stability of the modes for any value of wave number would be extremely complicated, we shall initially restrict our focus to the vanishing wave number limit, as before. In this limit, we have that the longitudinal modes related to net-baryon diffusion fluctuations can be cast in the form
\begin{equation}
\omega _{L,-}^{\mathrm{B}}=0,\left. {}\right. \omega _{L,+}^{\mathrm{B}%
}=\frac{i}{\gamma \left( \hat{\tau}_{n}-\hat{\tau}_{\kappa }V^{2}\right) },
\end{equation}%
while the remaining longitudinal modes become%
\begin{equation}
\omega _{\pm }^{\mathrm{sound}}=0,\left. {}\right. \omega _{0}^{\mathrm{shear}}=\frac{i\left( 3-V^{2}\right) }{\gamma \left[ 3\hat{\tau}_{\pi
}-\left( \hat{\tau}_{\pi }+4\right) V^{2}\right] }.
\end{equation}%

It is essential that the modes are stable at $k=0$. Furthermore, the perturbations must be stable for any value of the background fluid velocity. In order to ensure that the imaginary part of the mode has the correct sign (that leads to modes that are damped at late times) for any background velocity, we must have that,
\begin{equation}
\hat{\tau}_{n}\geq \hat{\tau}_{\kappa },\left. {}\right. \hat{\tau}_{\pi}\geq 2,\label{stab_cond}
\end{equation}%
which are the same conditions that are obtained when imposing causality for perturbations on top of a fluid that is at rest, see Eqs.~(\ref{causal02}) and (\ref{causal03}). These stability conditions are stronger than the ones found using the transverse modes and, therefore, are sufficient to ensure that the system is \textit{linearly} causal and stable. The condition for the shear relaxation time was previously obtained in \cite{rischke}. The stability condition that was obtained for the diffusion relaxation time is, to the best of our knowledge, new.

\begin{figure}[ht]
\begin{center}
\includegraphics[scale=0.35]{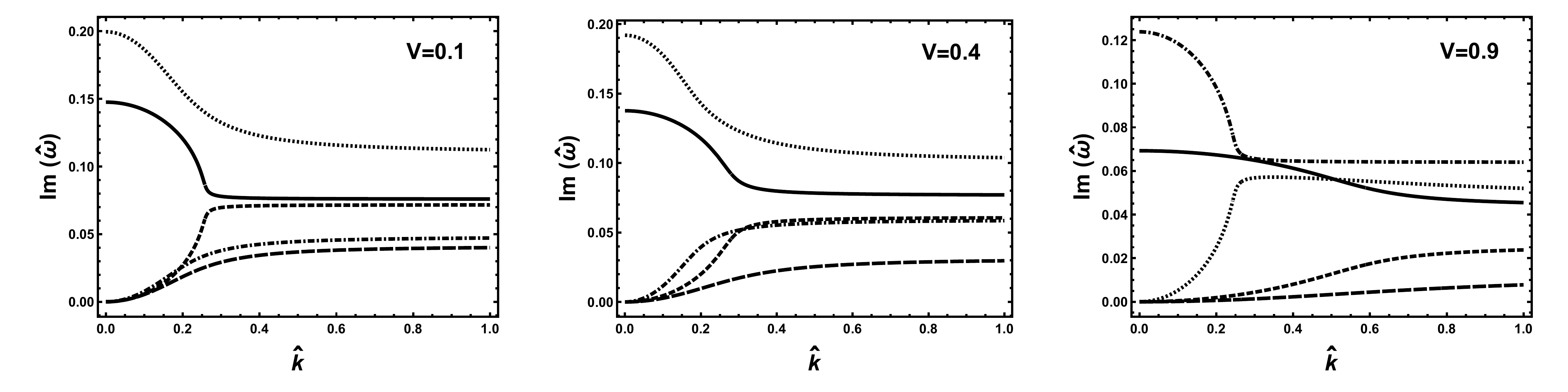} \\
\includegraphics[scale=0.35]{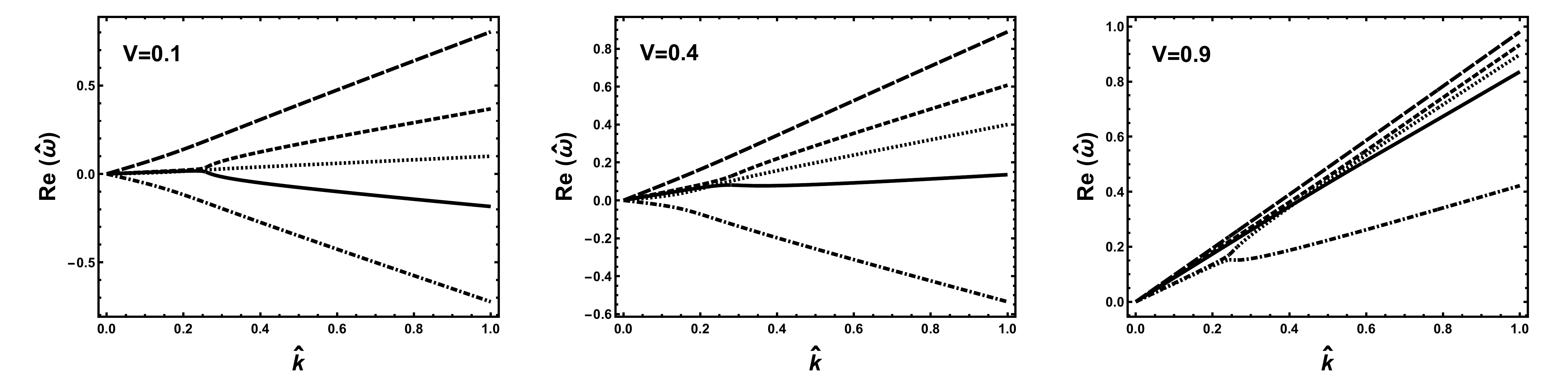}
\caption{The imaginary and real parts of the longitudinal modes for a moving background fluid, for $V=0.1$, $V=0.4$, and $V=0.9$ in the absence of coupling terms, i.e., $\hat{\mathcal{L}}_{n\pi}\hat{\mathcal{L}}_{n\pi}=0$.}
\label{long_wo_coup}
\end{center}
\end{figure}

\begin{figure}[ht]
\begin{center}
\includegraphics[scale=0.45]{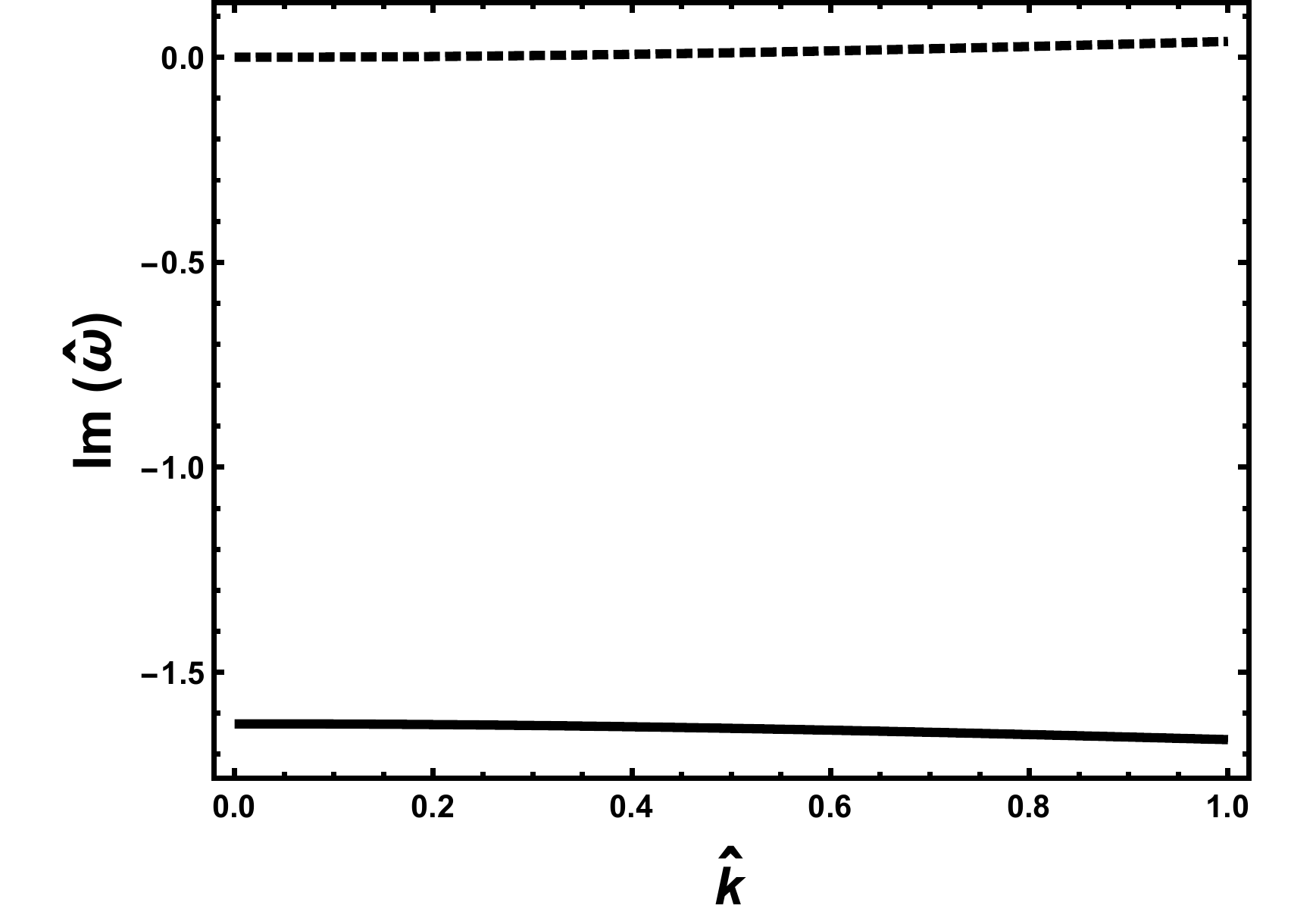}\hspace{1cm}\includegraphics[scale=0.45]{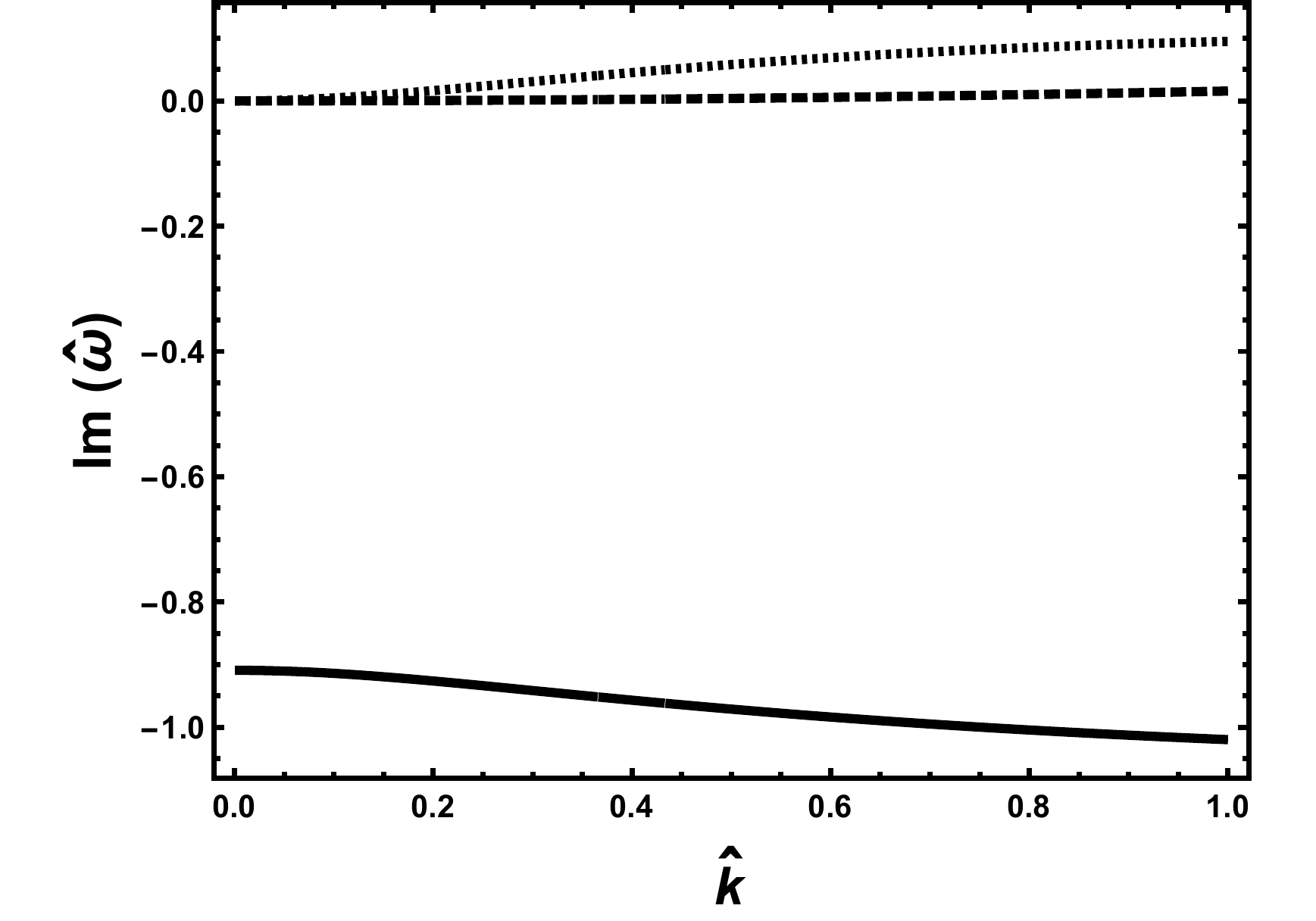}
\caption{Imaginary part of the unstable longitudinal mode related to baryon-number fluctuations for $\hat{\tau}_{\kappa}=9/16$ and $\hat{\tau}_{n}=3/16$ for $V=0.9$ (left panel) and imaginary part of the unstable shear mode for $\hat{\tau}_{\pi}=0.9$ for $V=0.9$ (right panel), proving the stability conditions we derived previously.}
\label{both_taus}
\end{center}
\end{figure}

The modes that are obtained as the solutions of Eq.~(\ref{disper02}) for a moving background fluid are displayed in Fig.~\ref{long_wo_coup}, considering $\hat\tau_\pi=5$, $\hat\tau_n=27/4$, and $\hat\tau_\kappa=9/16$. One can easily see that all modes are stable for all values of $\hat k$. Further, in Fig.~\ref{both_taus}, we consider two examples of unstable fluid configurations, i.e., fluids that do not satisfy the conditions derived in Eq.~(\ref{stab_cond}). On the left panel we analyze an unstable case in which $\hat{\tau}_{n} < \hat{\tau}_{\kappa}$. Here, we considered $\hat{\tau}_{n}=3/16$ and $\hat{\tau}_{\kappa}=9/16$ for an unperturbed system with velocity $V=0.9$. In this scenario, the longitudinal nonhydrodynamic mode related to net-baryon current fluctuations is unstable. On the right panel, we analyze the case where $\hat{\tau}_{\pi}=0.9$ for an unperturbed system with velocity $V=0.9$. Again, there is the occurrence of an unstable nonhydrodynamic mode, related to fluctuations of the shear-stress tensor.

So far, we have derived the stability conditions of the Israel-Stewart equations in the absence of coupling terms. These conditions, which are constraints for the relaxation times, were shown to be identical to the causality conditions obtained in the case where the unperturbed system is at rest. We concluded that the stability and causality conditions imposed by analyzing longitudinal modes are stronger and supersede the ones obtained investigating transverse modes. In the next section, we will be investigating the stability conditions in the presence of the coupling terms.

\section{Stability analysis with coupling terms}
\label{stab_with}

We now consider the case where coupling terms between the shear-stress tensor and diffusion 4-current are present. The following stability analysis will be very similar to the one performed in the previous section.

\subsection{Transverse modes}

As before, we start our discussion with the linearized equations of motion satisfied by the transverse degrees of freedom. The equations of motion can be cast in the following form
\begin{equation}
\left( 
\begin{array}{ccc}
\hat{\Omega} & -\hat{\kappa} & 0 \\ 
0 & i\mathcal{\hat{L}}_{n\pi }\hat{\kappa} & i\hat{\Omega}\hat{\tau}_{n}+1
\\ 
-i\hat{\kappa} & i\hat{\tau}_{\pi }\hat{\Omega}+1 & -i\frac{\mathcal{\hat{L}}%
_{\pi n}}{2}\hat{\kappa}%
\end{array}%
\right) \left( 
\begin{array}{c}
\delta \tilde{u}_{\bot }^{\mu } \\ 
\delta \tilde{\chi}_{\bot }^{\mu } \\ 
\delta \tilde{\xi}_{\bot }^{\mu }%
\end{array}%
\right) =0.
\end{equation}%
These equations lead to the following dispersion relation obtained from the determinant of the matrix above
\begin{equation}
-\mathcal{A}\hat{\Omega}^{3}+i\mathcal{B}\hat{\Omega}^{2}+\left( 1+\mathcal{C%
}\hat{\kappa}^{2}\right) \hat{\Omega}-i\hat{\kappa}^{2}=0,  \label{disper03}
\end{equation}%
where we defined the quantities%
\begin{eqnarray}
\mathcal{A} &\equiv &\hat{\tau}_{\pi }\hat{\tau}_{n},\left. {}\right. 
\mathcal{B}\equiv \hat{\tau}_{n}+\hat{\tau}_{\pi }, \label{def1}\\
\mathcal{C} &\equiv &\hat{\tau}_{n}-\frac{1}{2}\mathcal{\hat{L}}_{\pi n}%
\mathcal{\hat{L}}_{n\pi }.\label{def2}
\end{eqnarray}%
Naturally, we can see that the addition of coupling terms in the Israel-Stewart equations does not add a new mode to the dispersion relation for the transverse modes, since it continues to be a third-order polynomial. However, now that the coupling terms are included, the modes related to energy-momentum tensor fluctuations and net-baryon current fluctuations no longer
factorize. This renders the solution of the problem considerably more complicated and analytical solutions can no longer be cast in a simple form.
Furthermore, we also see that the modes only depend on the product of the
coupling terms, $\mathcal{\hat{L}}_{\pi n}\mathcal{\hat{L}}_{n\pi }$, that
is contained within our definitions of variables in the parameter $\mathcal{C}$.

Note that, if the relaxation times are set to zero, one no longer recovers a simple Navier-Stokes dispersion relation. Instead, one obtains the solution%
\begin{equation}
\hat{\Omega}=\frac{i\hat{\kappa}^{2}}{1-\frac{1}{2}\mathcal{\hat{L}}_{\pi n}%
\mathcal{\hat{L}}_{n\pi }\hat{\kappa}^{2}}.
\end{equation}%
This solution is purely imaginary and is only stable if $\mathcal{%
\hat{L}}_{\pi n}\mathcal{\hat{L}}_{n\pi }$ is negative. If the relaxation times are not zero, we will see that stable solutions with positive values of $\mathcal{\hat{L}}%
_{\pi n}\mathcal{\hat{L}}_{n\pi }$ are still possible, even though we will not investigate such cases thoroughly.   

\begin{figure}[ht]
\begin{center}
\includegraphics[scale=0.6]{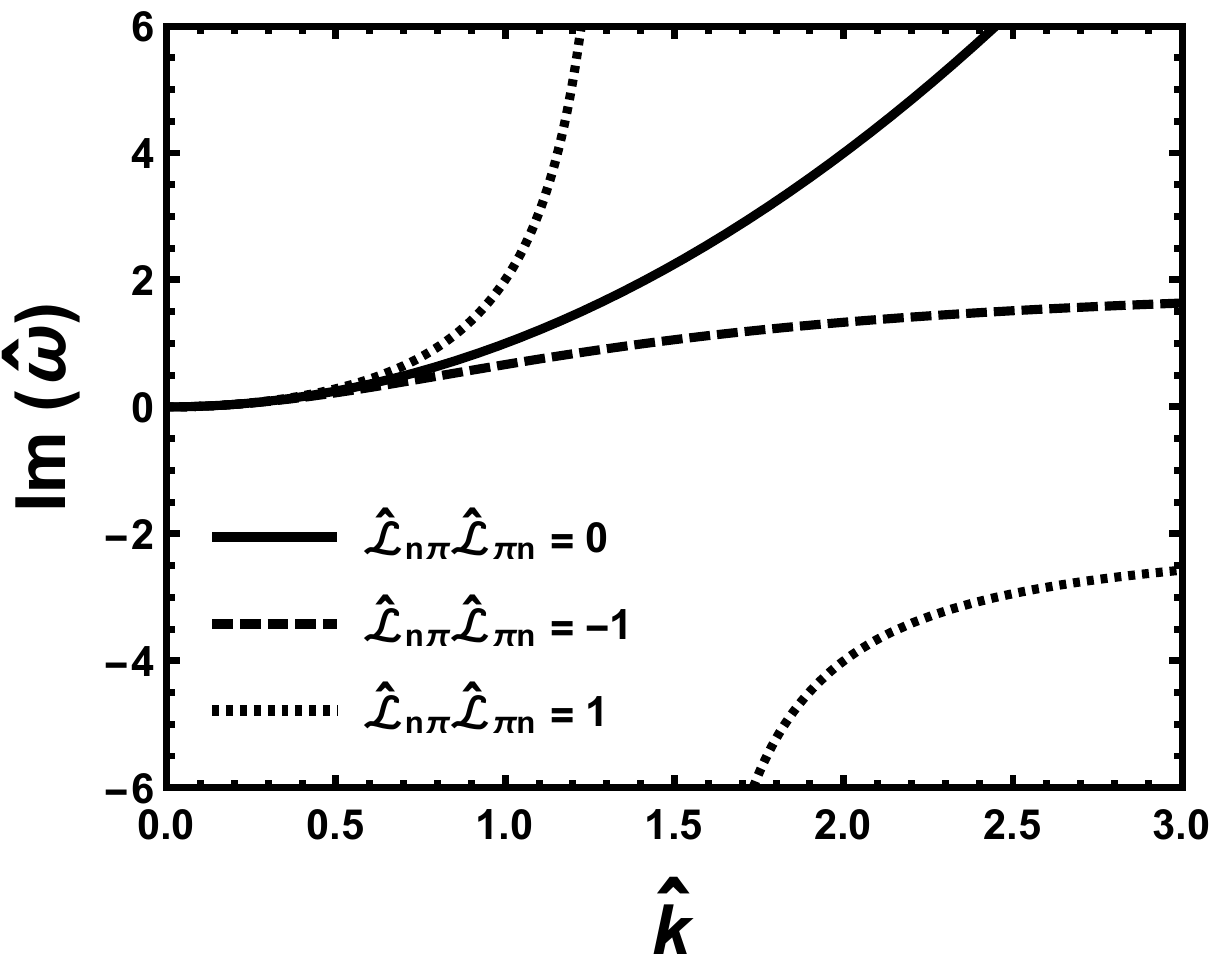}
\caption{The imaginary part of the dispersion relation of the transverse modes for an unperturbed system at rest, $V=0$, for three different values of the product of the coupling terms $\hat{\mathcal{L}}_{n\pi}\hat{\mathcal{L}}_{n\pi}=-1, 0, 1$. In this plot the relaxation times are set to zero, $\hat\tau_\pi=\hat\tau_n=0$.}
\label{transcoup1}
\end{center}
\end{figure}

We plot this mode assuming a static background ($\Omega = \omega$ and $\kappa = k$) in Fig.~\ref{transcoup1} for the following choices of coupling $\mathcal{\hat{L}}_{\pi n}\mathcal{\hat{L}}_{n\pi }=-1,$ $0,$
and $1$. Even though the modes show a familiar behavior in the small-wave-number region, as the wave number increases, they become rather different. Positive values of the effective coupling term render the theory unstable even in the case where the unperturbed system is at rest. On the other hand, for negative values, the imaginary part of the mode is positive definite, and therefore, it is stable. 

In the following stability analysis, all the relaxation times are once again taken into account. As before, we initially consider the case where the unperturbed system is at rest, $u^\mu_0=(1,0,0,0)$. In this case, the dispersion relation is

\begin{equation}
-\mathcal{A}\hat{\omega}^{3}+i\mathcal{B}\hat{\omega}^{2}+\left( 1+\mathcal{C%
}\hat{k}^{2}\right) \hat{\omega}-i\hat{k}^{2}=0.
\end{equation}%
We first study these solutions in two different limits: for small and large wave number. 

In the small-wave-number limit, $\hat{k}\ll 1$, one obtains
\begin{eqnarray}
\hat{\omega}_{T}^{\mathrm{diff}} &=&\frac{i}{\hat{\tau}_{n}}+\frac{i\mathcal{\hat{L}}_{\pi n}\mathcal{\hat{L}}_{n\pi }}{2(\hat{\tau}_\pi-\hat{\tau}_{n})}\hat{k}^{2}+\mathcal{O}\left( \hat{k}^{4}\right) , \\
\hat{\omega}_{T,+}^{\mathrm{shear}} &=&\frac{i}{\hat{\tau}_{\pi }}+\frac{i}{2%
}\frac{2\left( \hat{\tau}_{n}-\hat{\tau}_{\pi }\right) -\mathcal{\hat{L}}%
_{\pi n}\mathcal{\hat{L}}_{n\pi }}{\hat{\tau}_{\pi }-\hat{\tau}_{n}}\hat{k}%
^{2}+\mathcal{O}\left( \hat{k}^{4}\right) , \\
\hat{\omega}_{T,-}^{\mathrm{shear}} &=&i\hat{k}^{2}+\mathcal{O}\left( \hat{k}%
^{4}\right).
\end{eqnarray}
We see that, when either $\mathcal{\hat{L}}_{\pi n}$ or $\mathcal{\hat{L}}%
_{n\pi }$ is set to zero, we recover the result from the previous section. It is also interesting to notice that the coupling terms lead to corrections that are of higher order in $\hat{k}$. 

In the limit of large wave numbers, one can demonstrate that an expansion of the following form exists
\begin{equation}
\hat{\omega}=c_{-1}\hat{k}+\sum_{n=0}^{\infty }c_{n}\hat{k}^{-n}.
\end{equation}%
The expansion coefficients can be identified order by order by replacing the above ansatz into the dispersion relation. We then obtain the following solutions%
\begin{eqnarray}
\hat{\omega}_{T,\pm }^{\mathrm{shear}} &=&\pm \sqrt{\frac{\mathcal{C}}{%
\mathcal{A}}}\hat{k}+i\frac{\mathcal{BC}-\mathcal{A}}{2\mathcal{AC}}+O\left( 
\frac{1}{\hat{k}}\right) , \\
\hat{\omega}_{T}^{\mathrm{diff}} &=&\frac{i}{\mathcal{C}}+O\left( \frac{1}{%
\hat{k}}\right) .
\end{eqnarray}%
The solutions derived above, obtained from the asymptotic expansion of $\hat{\omega}$, already carry information on the linear stability of the theory. In order for the system to be stable, we must have that (i)$\sqrt{\mathcal{C}%
/\mathcal{A}}$ is real, (ii) $\mathcal{C}\geq 0$, and (iii) $\mathcal{BC}-%
\mathcal{A}\geq 0$. Note that condition (i) automatically guarantees
condition (ii), since $\mathcal{A}$ is guaranteed to be positive as long as the relaxation times are also positive. The stability conditions (ii) and (iii) lead to
\begin{eqnarray}
\mathcal{C} &>&0\Longrightarrow \mathcal{\hat{L}}_{\pi n}\mathcal{\hat{L}}%
_{n\pi }<2\hat{\tau}_{n}, \label{coup1}\\
\mathcal{BC}-\mathcal{A} &>&0\Longrightarrow \mathcal{\hat{L}}_{\pi n}%
\mathcal{\hat{L}}_{n\pi }<2\frac{\hat{\tau}_{n}^{2}}{\hat{\tau}_{n}+\hat{\tau%
}_{\pi }}. \label{coup2}
\end{eqnarray}
These conditions can be shown to be equivalent to those obtained using the Routh-Hurwitz stability criterion \cite{routh, hurwitz, kornbook}. In this sense, they are more general, and guarantee the stability conditions for any value of wave number $k$ (and not just for asymptotically large values of $k$). The last condition is clearly stronger and imposes restrictions on the values that the coupling terms coefficients can have. It is interesting that, with the inclusion of the coupling terms, we already have to impose stability conditions even for perturbations around a background that is at rest -- something that was not required in Navier-Stokes theory or in simplified versions of Israel-Stewart theory. Note that both conditions are automatically satisfied if the product of the coupling terms has a negative sign. Furthermore, a causality condition can be extracted from the expansion of the modes in the large-wave-number limit,
\begin{equation}
\lim_{k\rightarrow \infty}\left\vert\frac{\partial \mathrm{Re}(\omega)}{\partial k}\right\vert=\frac{\mathcal{C}}{\mathcal{A}}\leq 1\Longrightarrow 
\mathcal{\hat{L}}_{\pi n}\mathcal{\hat{L}}_{n\pi }\geq -2\hat{\tau}_{n}\left(\hat{\tau}_{\pi }-1\right). \label{causal_coup_trans}
\end{equation}

\begin{figure}[ht]
\begin{center}
\includegraphics[scale=0.35]{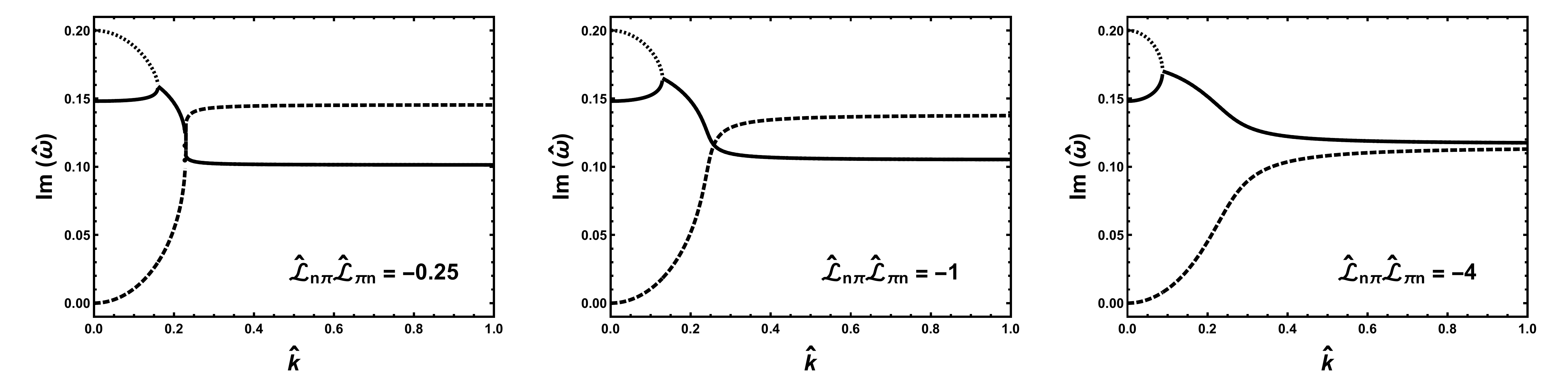}\\
\includegraphics[scale=0.35]{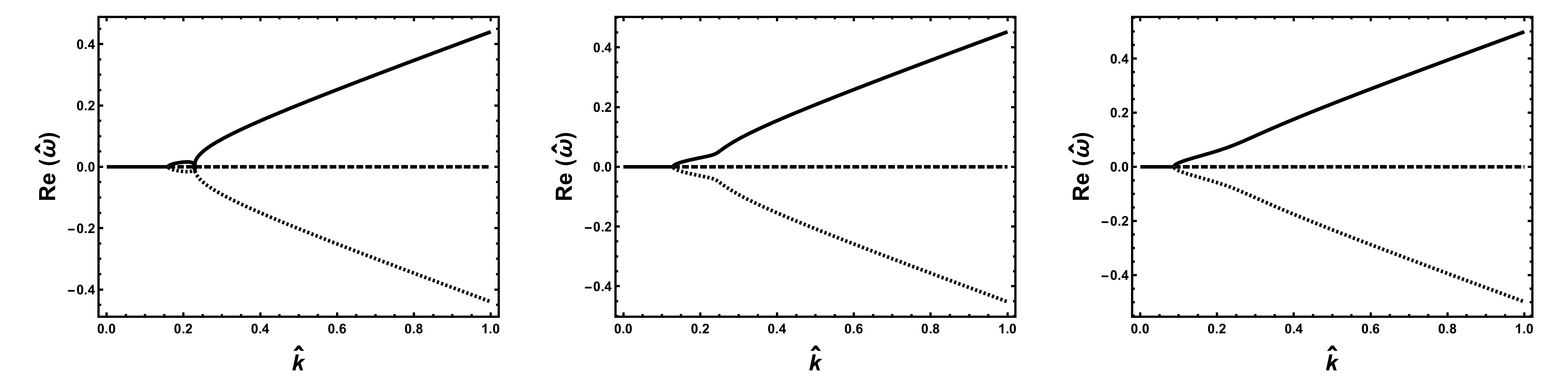}
\caption{Real and imaginary parts of the transverse modes for three negative values of the product of the coupling terms,  $\mathcal{\hat{L}}_{\pi n}\mathcal{\hat{L}}_{n\pi }=-0.25, -1, -4$, for a static background, $V=0$.}
\label{transnegcoups}
\end{center}
\end{figure}

\begin{figure}[ht]
\centering
\includegraphics[scale=0.35]{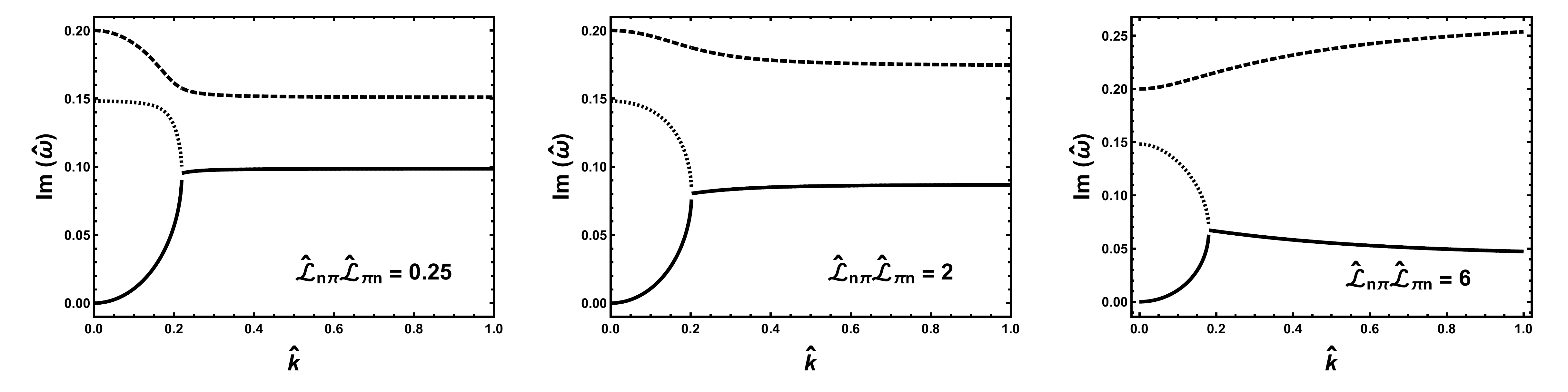}\\
\includegraphics[scale=0.35]{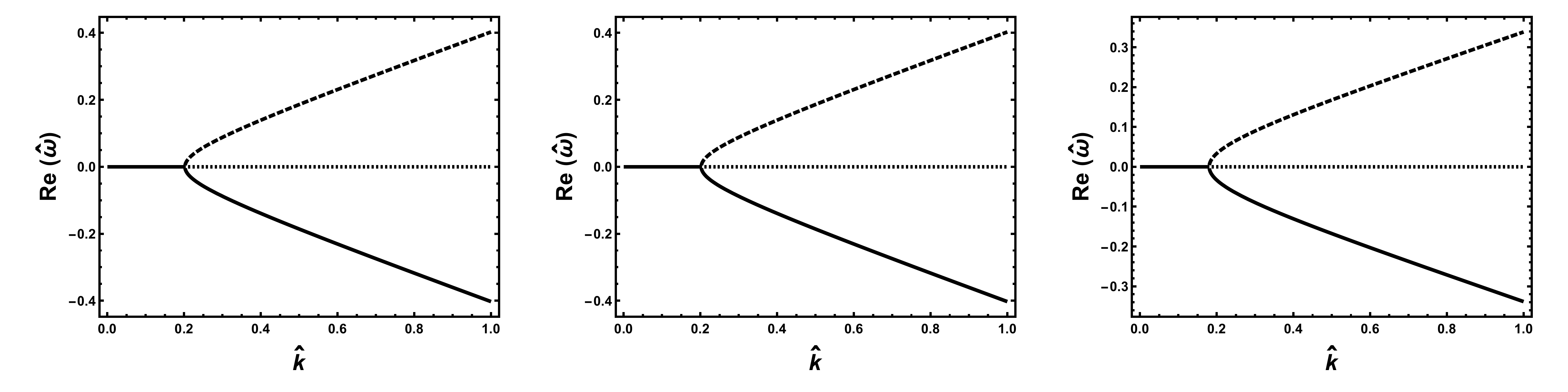}
\caption{Real and imaginary parts of the transverse modes for three positive values of the product of the coupling terms,  $\mathcal{\hat{L}}_{\pi n}\mathcal{\hat{L}}_{n\pi }=0.25, 2, 6$, for a static background, $V=0$.}
\label{transposcoups}
\end{figure}

\begin{figure}[ht]
\centering
\includegraphics[scale=0.45]{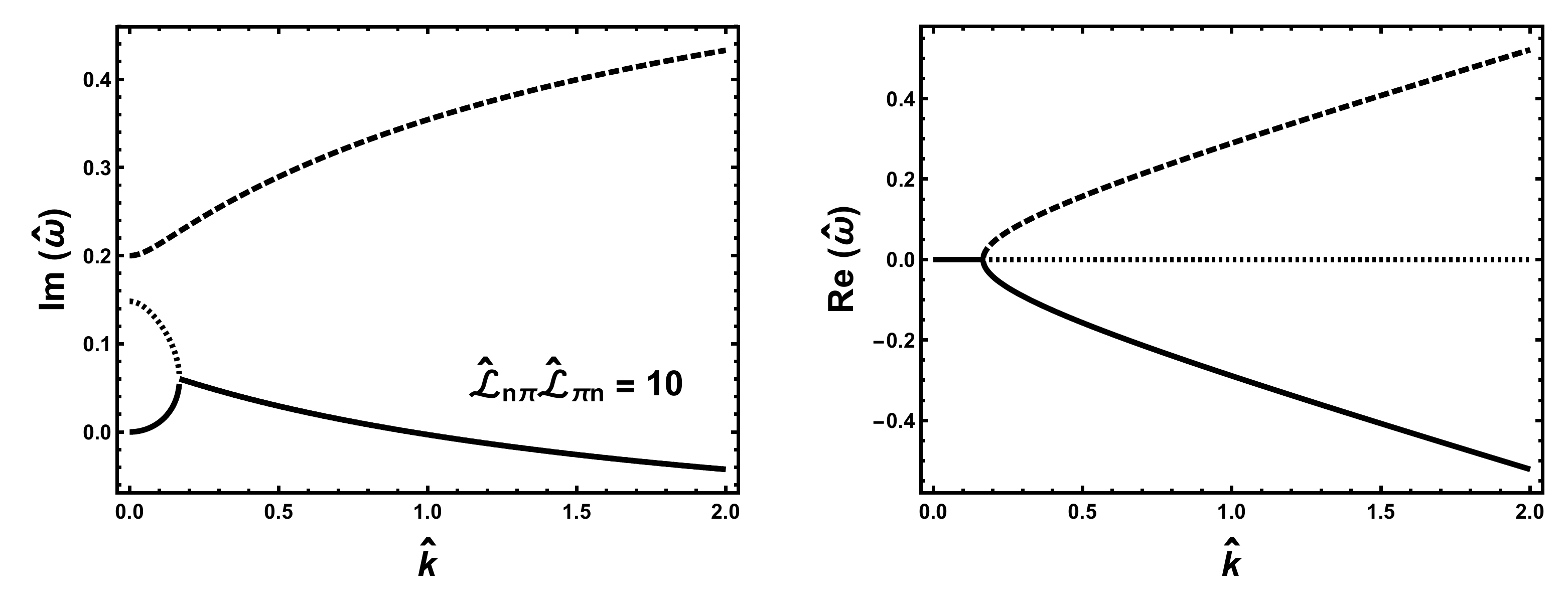}
\caption{Real and imaginary parts of the transverse modes for an unstable value for the product of the coupling terms, $\mathcal{\hat{L}}_{\pi n}\mathcal{\hat{L}}_{n\pi }=10$, for a static background, $V=0$.}
\label{transcoups}
\end{figure}

For the sake of illustration, we display, in Fig.~\ref{transnegcoups}, the real and imaginary parts of the modes $\hat{\omega}_{T,\pm}^{\mathrm{shear}}$ and $\hat{\omega}_{T}^{\mathrm{diff}}$ for the following negative values of the product of the coupling terms, $\mathcal{\hat{L}}_{\pi n}\mathcal{\hat{L}}_{n\pi }=-0.25, -1, -4$. In Fig.~\ref{transposcoups}, we display such modes for positive values of $\mathcal{\hat{L}}_{\pi n}\mathcal{\hat{L}}_{n\pi }$ chosen to be $\mathcal{\hat{L}}_{\pi n}\mathcal{\hat{L}}_{n\pi }=0.25, 2, 6$. As before, these plots were made using the transport coefficients calculated from the Boltzmann equation,  $\hat{\tau}_{\pi }=5$
and $\hat{\tau}_{n}=27/4$. For these values of transport coefficients, the stability condition given by Eq.~(\ref{coup2}) becomes, $\mathcal{\hat{L}}_{\pi n}\mathcal{\hat{L}}_{n\pi }\lesssim 7.75$. We note that it is possible to obtain stable modes even for positive values of $\mathcal{\hat{L}}_{\pi n}\mathcal{\hat{L}}_{n\pi }$. As the value of $\mathcal{\hat{L}}_{\pi n}\mathcal{\hat{L}}_{n\pi }$ becomes negative, the coupling terms render the imaginary parts of the nonhydrodynamic modes degenerate at larger values of wave numbers. On the other hand, if $\mathcal{\hat{L}}_{\pi n}\mathcal{\hat{L}}_{n\pi }$ is positive, the nonhydrodynamic mode related to the diffusion 4-current, $\hat{\omega}_{T}^{\mathrm{diff}}$, becomes degenerate with the hydrodynamic mode $\hat{\omega}_{T,-}^{\mathrm{shear}}$ instead, when the wave number increases. In Fig.~\ref{transcoups}, we show a case in which the relaxation times are chosen in such a way to ensure the causality and stability conditions derived in the previous sections, but the modes are driven unstable by the coupling term, with $\mathcal{\hat{L}}_{\pi n}\mathcal{\hat{L}}_{n\pi }=10$.

\begin{figure}[ht]
\begin{center}
\includegraphics[scale=0.3]{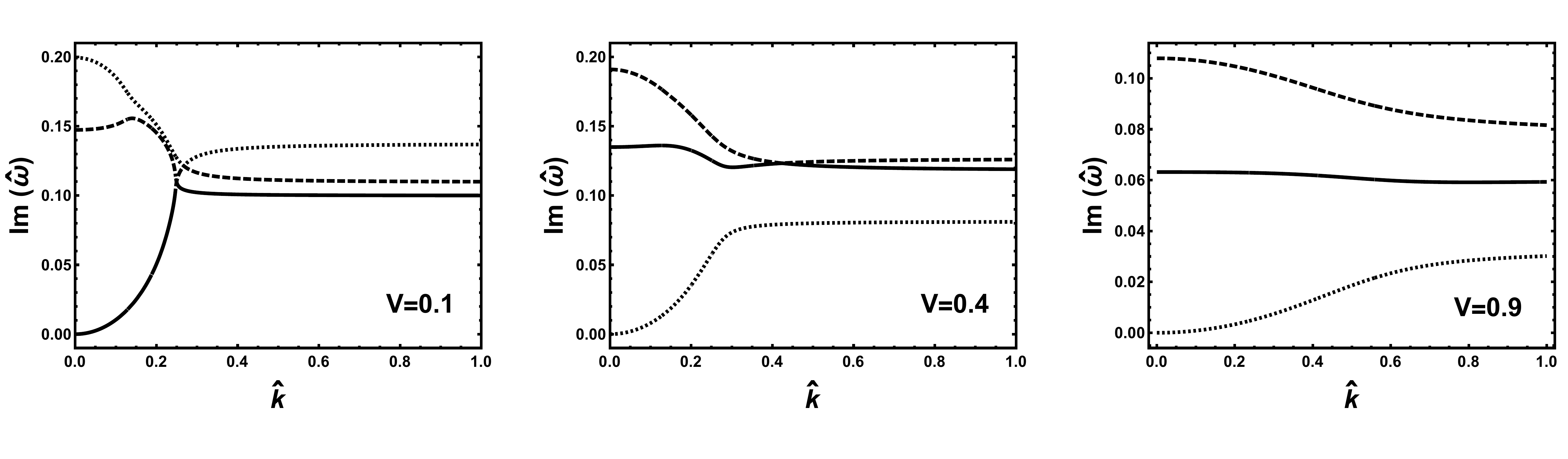}\\
\includegraphics[scale=0.3]{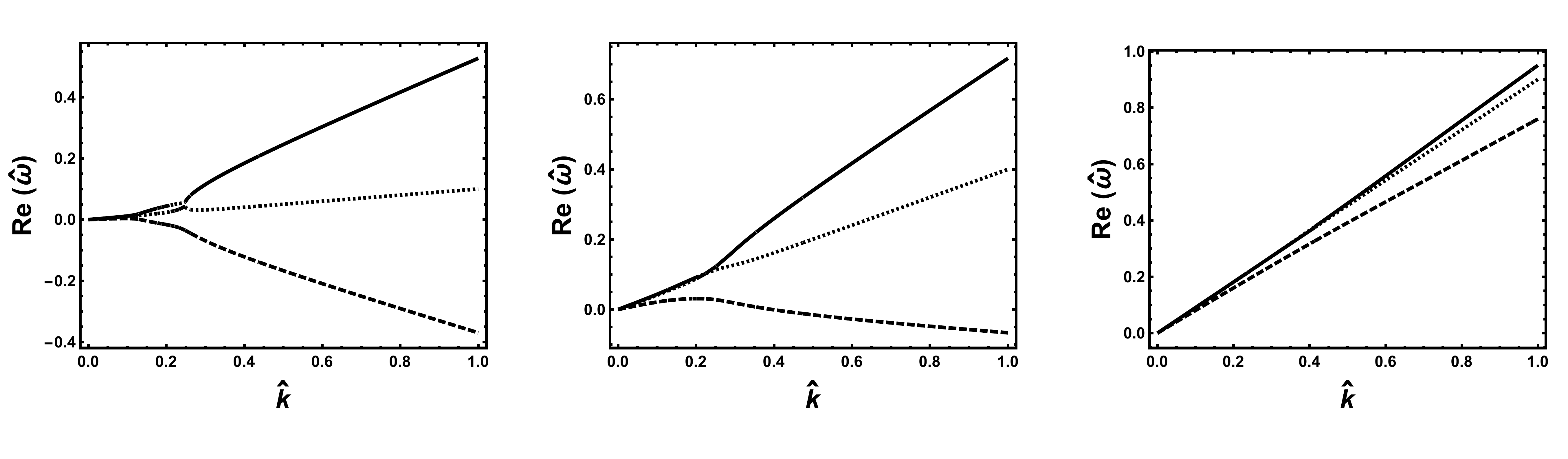}
\caption{Real and imaginary parts of the transverse modes considering a negative value for the product of the coupling terms, $\mathcal{\hat{L}}_{\pi
n}\mathcal{\hat{L}}_{n\pi }=-1$, for three different values of the background velocity $V=0.1$, $V=0.4$, and $V=0.9$.}
\label{transboostedneg}
\end{center}
\end{figure}

\begin{figure}[ht]
\begin{center}
\includegraphics[scale=0.35]{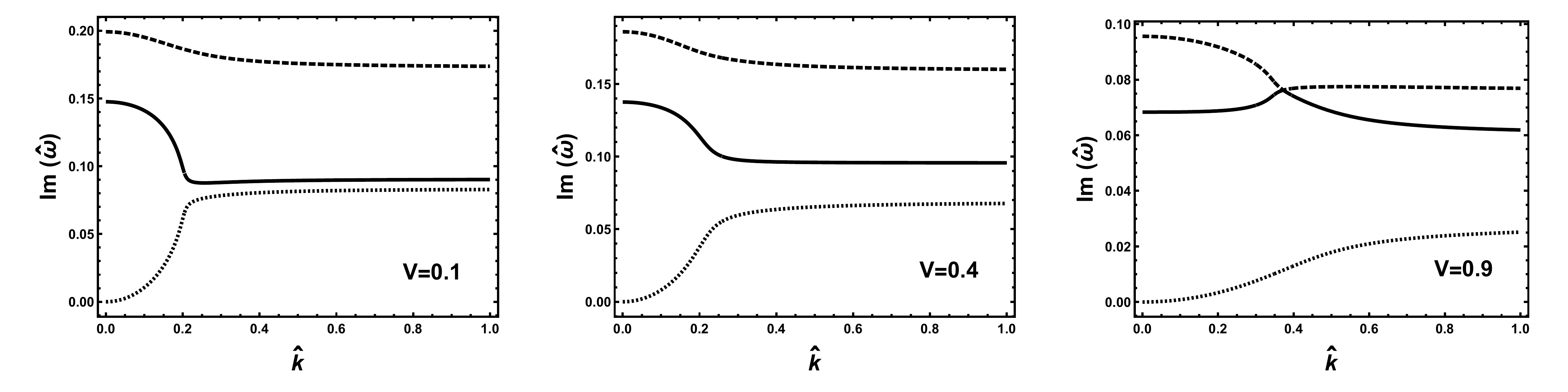}\\
\includegraphics[scale=0.4]{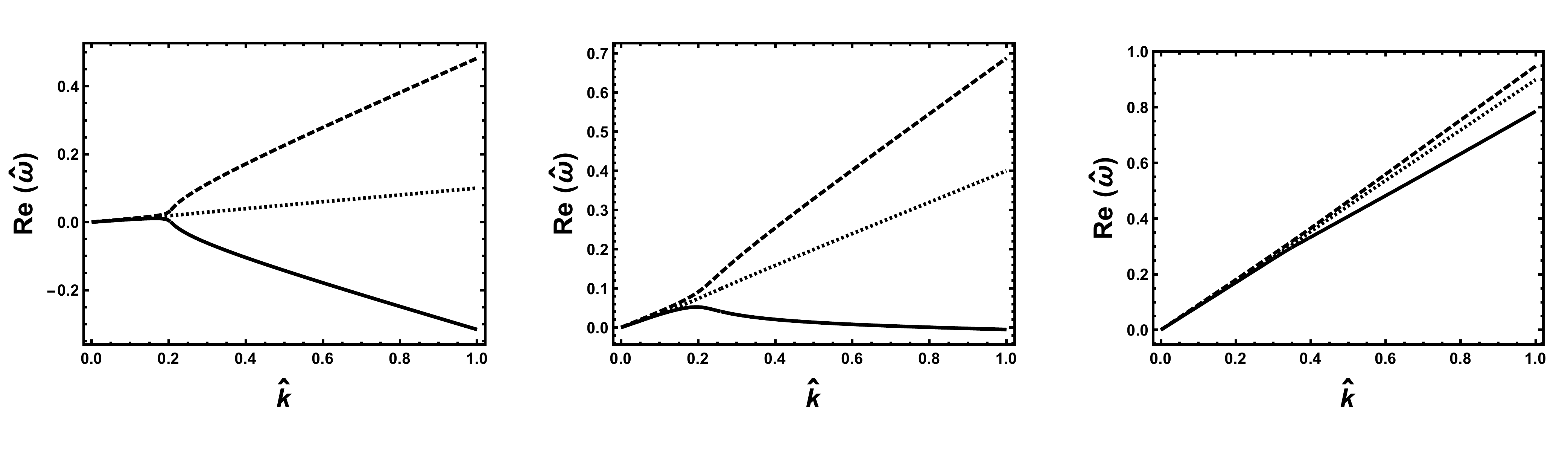}
\caption{Real and imaginary parts of the transverse modes considering a positive value for the product of the coupling terms, $\mathcal{\hat{L}}_{\pi
n}\mathcal{\hat{L}}_{n\pi }=2$, for three different values of the background velocity $V=0.1$, $V=0.4$, and $V=0.9$.}
\label{transboostedpos}
\end{center}
\end{figure}

We now perform the same analysis considering that the unperturbed fluid is moving with a velocity $V$ in the, e.g. $x$ direction. Once again, we assume that the 4-velocity and the perturbations are taken in the same direction, so that $u_{0}^{\mu }=\gamma(1,V,0,0)
$ and $k^{\mu }=(\omega ,k,0,0)$. As already discussed, this leads to Eqs.~(\ref{omega}) and (\ref{kappa}), which must be inserted into Eq.~(\ref{disper03}) in order to obtain the dispersion relation satisfied by $\omega$ and $k$. For the sake of illustration, the solutions of the resulting dispersion relation are plotted for two different cases: in Fig.~\ref{transboostedneg}, for negative values of $\mathcal{\hat{L}}_{\pi n}\mathcal{\hat{L}}_{n\pi }$, and in Fig.~\ref{transboostedpos}, for positive values of $\mathcal{\hat{L}}_{\pi n}\mathcal{\hat{L}}_{n\pi }$ considering the following values of the background fluid velocity, $V=0.1$, $V=0.4$, and $V=0.9$.

Similar to what was done in the previous sections, we discuss the linear stability of the theory by analyzing the modes at zero wave number, $\hat{k}=0$. This simplifies considerably the calculations and allows us to provide basic necessary conditions for linear stability. In this limit, the dispersion relation reduces to
\begin{equation}
-\mathcal{A}\left( \gamma \hat{\omega}\right) ^{3}+i\mathcal{B}\left( \gamma 
\hat{\omega}\right) ^{2}+\left[ 1+\mathcal{C}\left( \gamma \hat{\omega}%
V\right) ^{2}\right] \left( \gamma \hat{\omega}\right) -i\left( \gamma \hat{%
\omega}V\right) ^{2}=0.  \notag
\end{equation}%
Thus, it is possible to express the solutions in a simple analytical form
\begin{equation}
\hat{\omega}=0,\gamma \hat{\omega}_{\pm }=i\frac{\left( \mathcal{B}%
-V^{2}\right) \pm \sqrt{\left( \mathcal{B}-V^{2}\right) ^{2}-4\left( 
\mathcal{A}-\mathcal{C}V^{2}\right) }}{\mathcal{A}-\mathcal{C}V^{2}}. \label{trans_k_zero}
\end{equation}%
Naturally, the vanishing solution corresponds to the hydrodynamic mode, while the remaining solution corresponds to the nonhydrodynamic mode. The stability of the perturbations is usually governed by the nonhydrodynamic mode, which will either decay or increase exponentially depending on the choices of transport coefficients. If $V=0$, we recover the previous solutions, $\hat{\omega}_{\pm}=i/\hat{\tau}_{\pi},i/\hat{\tau}_{n}$.

Next, we investigate the necessary conditions the transport coefficients must satisfy in order for the nonhydrodynamic modes $\hat{\omega}_\pm$ in Eq.~(\ref{trans_k_zero}) to be both stable in the homogeneous limit. Naturally, in order for the modes to have a positive imaginary part at $k=0$, the numerator and denominator must carry the same sign. The denominator should not change sign as the velocity of the background fluid increases, otherwise resulting in an instability. Therefore, stable fluid-dynamical formulations must satisfy $\mathcal{A}>\mathcal{C}V^2$ for all values of $0\leq V\leq1$, which is guaranteed by the following condition 
\begin{equation}
\mathcal{A}>\mathcal{C}. \label{a>c}
\end{equation}
The imaginary part of the numerator must also be positive. This condition requires, at least, that $\mathcal{B}-V^2>0$ for all values of $0\leq V\leq1$, leading to
\begin{equation}
\mathcal{B}>1. \label{b>1}
\end{equation}
Finally, we study the term inside the square root in Eq.~(\ref{trans_k_zero}). One can show that for negative values of the product of the coupling terms, $\hat{\mathcal{L}}_{n\pi}\hat{\mathcal{L}}_{\pi n}\leq0$, which is the case considered here, such term is positive and smaller than $\mathcal{B}-V^2$, and does not affect the sign of the modes. Furthermore, considering also positive values for the product between the coupling terms, $\hat{\mathcal{L}}_{n\pi}\hat{\mathcal{L}}_{\pi n}\geq0$, either the term inside the square root remains positive and smaller than $\mathcal{B}-V^2$ or it becomes negative, resulting in an imaginary value for the square root. Either way, the conditions given by Eqs.~(\ref{a>c}) and (\ref{b>1}) are sufficient to maintain the theory linearly stable at $k=0$. The conditions (\ref{a>c}) and (\ref{b>1}) can be expressed as the following constraints for the transport coefficients
\begin{eqnarray}
\mathcal{B} &>&1\Longrightarrow \hat{\tau}_{n}+\hat{\tau}_{\pi }>1, \\
\mathcal{A} &>&\mathcal{C}\Longrightarrow \mathcal{\hat{L}}_{\pi n}\mathcal{%
\hat{L}}_{n\pi }\geq -2\hat{\tau}_{n}\left( \hat{\tau}_{\pi }-1\right) .
\end{eqnarray}
The latter is equivalent to the causality condition obtained for
perturbations around a background at rest, see Eq.~(\ref{causal_coup_trans}).
Combining the stability conditions above with those derived for a background at rest, see Eqs.~(\ref{coup1}) and (\ref{coup2}), we then obtain
\begin{eqnarray}
\hat{\tau}_{n}+\hat{\tau}_{\pi } &>&1, \\
-2\hat{\tau}_{n}\left( \hat{\tau}_{\pi }-1\right) &\leq &\mathcal{\hat{L}}
_{\pi n}\mathcal{\hat{L}}_{n\pi }<\frac{2\hat{\tau}_{n}^{2}}{\hat{\tau}_{n}+
\hat{\tau}_{\pi }}.
\end{eqnarray}%
These inequalities can be further simplified by imposing that the product between the coupling terms is negative, $\mathcal{\hat{L}}_{\pi n}\mathcal{\hat{L}}_{n\pi }\leq 0$. In this case, the last relation reduces to
\begin{equation}
|\mathcal{\hat{L}}_{\pi n}\mathcal{\hat{L}}_{n\pi }|
\leq 2\hat{\tau}_{n}\left( \hat{\tau}_{\pi }-1\right),\text{    }\hat{\tau}_{\pi }\geq 1,\text{    }\hat{\tau}_{n}\geq 0
\end{equation}
with the conditions related to the relaxation times being equivalent to the stability conditions obtained in the absence of coupling terms. However, we note that positive values of $\mathcal{\hat{L}}_{\pi n}\mathcal{\hat{L}}_{n\pi }$ can also lead to stable theories. In this case, it would even be possible to violate the condition $\hat{\tau}_{\pi}\geq 1$. As a matter of fact, if $\hat{\tau}_{\pi }\leq 1$, we see that only positive values of $\mathcal{\hat{L}}_{\pi n}\mathcal{\hat{L}}_{n\pi}$ are allowed. 

Then, removing the scaling factors from the transport coefficients, we obtain the following condition
\begin{equation}
\left\vert \ell _{\pi n}\ell _{n\pi }\right\vert \leq 2\tau _{n}\left( \tau
_{\pi }-\tau _{\eta }\right) . \label{stab_trans_olson}
\end{equation} 
A similar condition can be recovered from Olson's original work \cite{olson} by imposing that the transverse characteristic velocities are subluminal [Eq.~(91) of the aforementioned paper]. This condition is satisfied in calculations from the Boltzmann equation
\cite{dnmr, 14moment}. So far, we are not aware of any microscopic calculations that do not satisfy this condition.

\subsection*{Longitudinal modes}

In the presence of coupling terms, the linearized equations for the longitudinal modes
can be expressed in the following form
\begin{equation}
\left( 
\begin{array}{ccccc}
\Omega & 0 & 0 & -\kappa & 0 \\ 
0 & \Omega & -\kappa & 0 & 0 \\ 
0 & -\frac{\kappa}{3} & \Omega & 0 & -\kappa \\ 
-i\hat{\tau}_{\kappa }\hat{\kappa} & 0 & 0 & i\hat{\tau}_{n}\hat{\Omega}+1 & 
i\mathcal{\hat{L}}_{n\pi }\hat{\kappa} \\ 
0 & 0 & -\frac{4}{3}i\hat{\kappa} & -\frac{2}{3}i\mathcal{\hat{L}}_{\pi n}%
\hat{\kappa} & i\hat{\tau}_{\pi }\hat{\Omega}+1%
\end{array}%
\right) \left( 
\begin{array}{c}
\delta \tilde{n}_{B}/n_{0} \\ 
\delta \tilde{\varepsilon}/w_{0} \\ 
\delta \tilde{u}_{\Vert } \\ 
\delta \tilde{\xi}_{\Vert } \\ 
\delta \tilde{\chi}_{\Vert }%
\end{array}%
\right) =0.
\end{equation}%
We then obtain the following dispersion relation%
\begin{equation}
\left[ \left( \hat{\Omega}^{2}-\frac{1}{3}\hat{\kappa}^{2}\right) (i\hat{\tau%
}_{\pi }\hat{\Omega}+1)-\frac{4}{3}i\hat{\kappa}^{2}\hat{\Omega}\right] %
\left[ \hat{\Omega}(i\hat{\tau}_{n}\hat{\Omega}+1)-i\hat{\tau}_{\kappa }\hat{%
\kappa}^{2}\right] -\frac{2}{3}\mathcal{\hat{L}}_{\pi n}\mathcal{\hat{L}}%
_{n\pi }\left( \hat{\Omega}^{2}-\frac{1}{3}\hat{\kappa}^{2}\right) \hat{%
\Omega}\hat{\kappa}^{2}=0. \label{disper04}
\end{equation}%
For the sake of convenience, we rewrite this expression as%
\begin{equation}
-\mathcal{A}\hat{\Omega}^{5}+i\mathcal{B}\hat{\Omega}^{4}+\left( 1+2\mathcal{%
AS}\hat{\kappa}^{2}\right) \hat{\Omega}^{3}-i\frac{\mathcal{BD}}{3}\hat{%
\kappa}^{2}\hat{\Omega}^{2}-\frac{1}{3}\left( 1+\mathcal{E}\hat{\kappa}%
^{2}\right) \hat{\Omega}\hat{\kappa}^{2}+i\frac{\hat{\tau}_{\kappa }}{3}\hat{%
\kappa}^{4}=0,
\end{equation}%
where $\mathcal{A}$, $\mathcal{B}$, and $\mathcal{C}$ were defined in the
previous section, in Eqs.~(\ref{def1}) and (\ref{def2}), and we further introduced the variables
\begin{eqnarray}
\mathcal{S} &\equiv &\frac{\mathcal{A}+3\hat{\tau}_{\pi }\hat{\tau}_{\kappa
}+4\mathcal{C}}{6\mathcal{A}}, \\
\mathcal{D} &\equiv &\frac{\mathcal{B}+3\hat{\tau}_{\kappa }+4}{\mathcal{B}},
\\
\mathcal{E} &\equiv &4\hat{\tau}_{\kappa }+\hat{\tau}_{\pi }\hat{\tau}%
_{\kappa }+\frac{4}{3}\left( \mathcal{C}-\hat{\tau}_{n}\right) .
\end{eqnarray}
Other useful definitions that will be employed in the remaining of this
paper are
\begin{equation}
\mathcal{M}\equiv \frac{\mathcal{E}}{3\mathcal{A}},\left. {}\right. \mathcal{%
R}\equiv \sqrt{\mathcal{S}^{2}-\mathcal{M}}.
\end{equation}

Initially, we shall consider the case where the unperturbed system is at
rest, i.e., $u_{0}^{\mu }=\left( 1,0,0,0\right) $. In this case, the
dispersion relation is%
\begin{equation}
-\mathcal{A}\hat{\omega}^{5}+i\mathcal{B}\hat{\omega}^{4}+\left( 1+2\mathcal{%
AS}\hat{k}^{2}\right) \hat{\omega}^{3}-i\mathcal{BD}\hat{k}^{2}\hat{\omega}%
^{2}-\frac{1}{3}\left( 1+\mathcal{E}\hat{k}^{2}\right) \hat{\omega}\hat{k}%
^{2}+i\frac{\hat{\tau}_{\kappa }}{3}\hat{k}^{4}=0.
\end{equation}%
The solutions of this polynomial equation are rather complicated. As before, we shall look at such solutions in the limits of small and large wave number in order to extract some information about the behavior of the modes. In this case, for small wave numbers, $\hat{k}\ll 1$, we have that
\begin{eqnarray}
\omega _{L,+}^{\mathrm{B}} &=&\frac{i}{\hat{\tau}_{n}}+\mathcal{O}\left( 
\hat{k}^{2}\right) , \\
\omega _{L}^{\mathrm{shear}} &=&\frac{i}{\hat{\tau}_{\pi }}+\mathcal{O}%
\left( \hat{k}^{2}\right) , \\
\omega _{L,-}^{\mathrm{B}} &=&i\hat{\tau}_{\kappa }\hat{k}^{2}+\mathcal{O}%
\left( \hat{k}^{3}\right) , \\
\omega _{\pm }^{\mathrm{sound}} &=&\pm \frac{1}{\sqrt{3}}\hat{k}+\frac{2}{3%
}i\hat{k}^{2}+\mathcal{O}\left( \hat{k}^{3}\right) .
\end{eqnarray}%
We have two nonhydrodynamic modes and three hydrodynamic modes. At small
wave numbers, the hydrodynamic modes behave like the modes found in Navier-Stokes theory. Furthermore, for large wave numbers, the modes can be written as 
\begin{eqnarray}
\hat{\omega} &=&\pm\hat{k}\sqrt{\mathcal{S}\pm \mathcal{R}}+\frac{i}{%
\mathcal{A}}\frac{3\mathcal{B}\left( \mathcal{S}\pm \mathcal{R}\right) ^{2}-%
\mathcal{BD}\left( \mathcal{S}\pm \mathcal{R}\right) +\hat{\tau}_{\kappa }}{%
15\left( \mathcal{S}\pm \mathcal{R}\right) ^{2}-18\mathcal{S}\left( \mathcal{%
S}\pm \mathcal{R}\right) +3\mathcal{M}}+\mathcal{O}\left( \frac{1}{\hat{k}}%
\right) , \label{longeq1}\\
\hat{\omega} &=&i\frac{\hat{\tau}_{\kappa }}{\mathcal{E}}+\mathcal{O}\left( 
\frac{1}{\hat{k}}\right).
\end{eqnarray}%
In order for these modes to be stable, it is required that their imaginary parts are positive and thus we have the following necessary conditions:
\begin{enumerate}[label=(\roman*)]
\item $\hat{\tau}_{\kappa }/\mathcal{E}>0$; \label{i}
\item $\sqrt{\mathcal{S}\pm \mathcal{R}}$ is real; \label{ii}
\item $\left[ 3\mathcal{B}\left( \mathcal{S}\pm \mathcal{R}\right) ^{2}-\mathcal{BD}\left( \mathcal{S}\pm \mathcal{R}\right) +\hat{\tau}
_{\kappa }\right] /\left[ 15\left( \mathcal{S}\pm \mathcal{R}\right) ^{2}-18\mathcal{S}\left( \mathcal{S}\pm \mathcal{R}\right) +3\mathcal{M}\right]$ is positive. \label{iii}
\end{enumerate}
The causality condition further imposes the following constraints for the asymptotic group velocity 
\begin{equation}
\lim_{k\rightarrow \infty}\left\vert\frac{\partial \mathrm{Re}(\omega)}{\partial k}\right\vert=\sqrt{\mathcal{S}\pm \mathcal{R}}\leq1. \label{causality_coup_long}
\end{equation}
In the following, we translate the conditions above into conditions for the transport coefficients. 

Condition \ref{i} is guaranteed as long as $\mathcal{E}>0$, which leads to another constraint for the product of the coupling coupling terms, given by
\begin{equation}
\mathcal{\hat{L}}_{\pi n}\mathcal{\hat{L}}_{n\pi}<\frac{3}{2}\hat{\tau}_{\kappa}\left(\hat{\tau}_{\pi }+4\right),
\end{equation}
which is automatically satisfied if $\mathcal{\hat{L}}_{\pi n}\mathcal{\hat{L}}_{n\pi}<0$. Condition \ref{ii} is satisfied if $\mathcal{S}\geq\mathcal{R}$ (this condition is automatically satisfied if the product of the coupling terms is negative) and if $\mathcal{R}$ is real. The latter condition further implies that
\begin{equation}
\mathcal{S}^2>\mathcal{M}.
\end{equation}
This condition will be further developed when we discuss the modes for perturbations on top of a moving background fluid. Finally, condition \ref{iii} is satisfied as long as
\begin{eqnarray}
3\mathcal{B}\left( \mathcal{S}+\mathcal{R}\right) ^{2}-\mathcal{BD}
\left( \mathcal{S}+\mathcal{R}\right) +\hat{\tau}_{\kappa }>0,\label{s+r}\\
3\mathcal{B}\left( \mathcal{S}-\mathcal{R}\right) ^{2}-\mathcal{BD}
\left( \mathcal{S}-\mathcal{R}\right) +\hat{\tau}_{\kappa }<0.\label{s-r}
\end{eqnarray}
If the product of the coupling terms is negative, these conditions are automatically satisfied and do not provide any new constraints for the stability of the system. On the other hand, if the product of the coupling terms is positive, these conditions will lead to constraints for the transport coefficients, but we shall not discuss or explicitly derive them in this work. Furthermore, all conditions listed above can be shown to be equivalent to the ones found by using the Routh-Hurwitz criterion \cite{routh,hurwitz,kornbook}. In this sense, they are valid not only in the large-wave-number regime, but also for any value of wave number $k$.
\begin{figure}[ht]
\begin{center}
\includegraphics[scale=0.47]{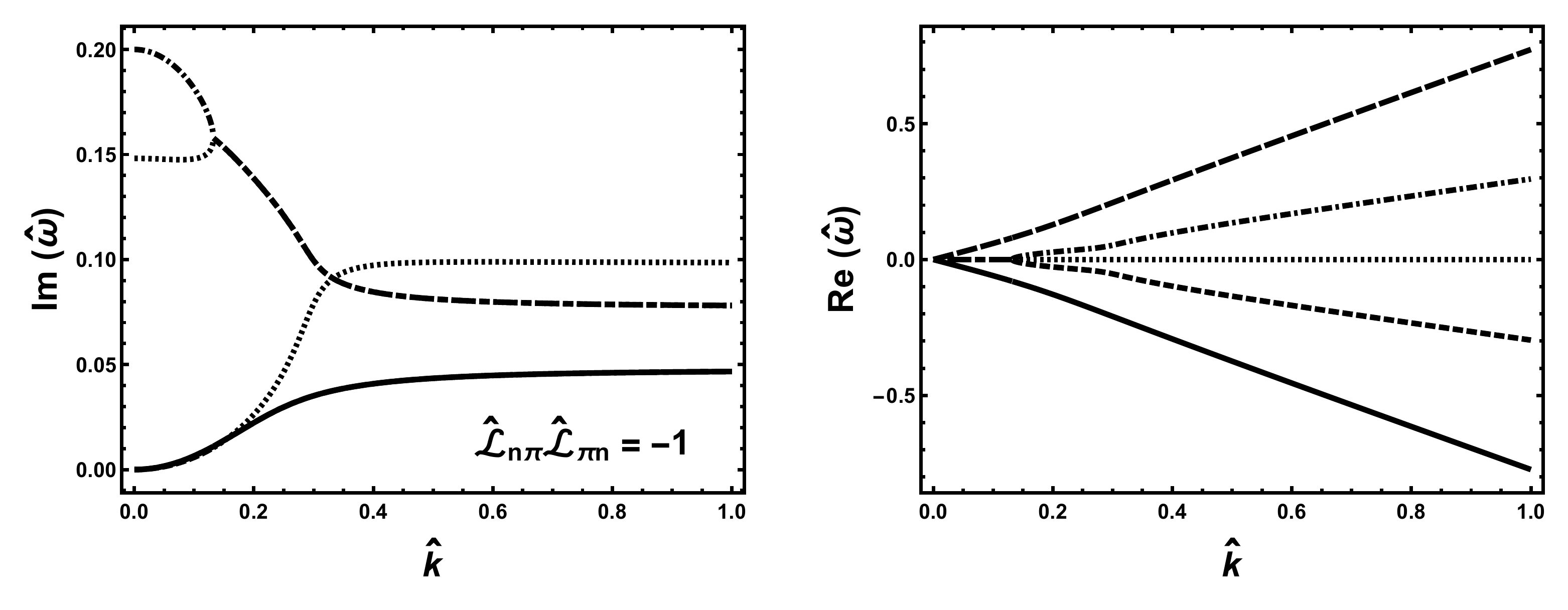}\\
\includegraphics[scale=0.47]{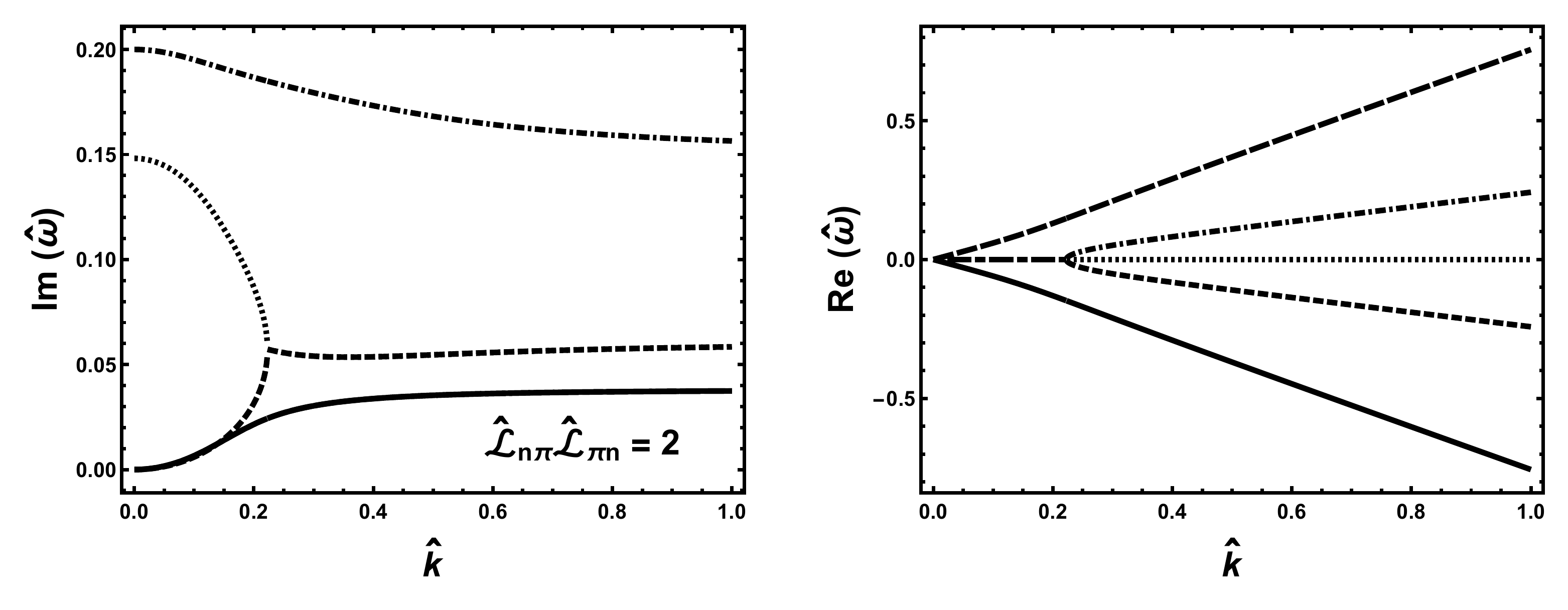}
\caption{Real and imaginary parts of the longitudinal modes considering a negative and a positive values for the product of the coupling terms, $\mathcal{\hat{L}}_{\pi
n}\mathcal{\hat{L}}_{n\pi }=-1, 2$ for a background at rest, $V=0$.}
\label{longrestcoup0}
\end{center}
\end{figure}

The solutions of Eq.~(\ref{disper04}) for a static background are displayed in Fig.~\ref{longrestcoup0} for a negative value of the coupling term $\mathcal{\hat{L}}_{\pi n}\mathcal{\hat{L}}_{n\pi }=-1$ (top panels), and also for a positive value $\mathcal{\hat{L}}_{\pi n}\mathcal{\hat{L}}_{n\pi }=2$ (bottom panels). The inclusion of the coupling produces a similar behavior when compared to the transverse modes, see Figs.~\ref{transnegcoups} and \ref{transposcoups}, where the sign of the product of the coupling terms dictates which modes merge at large values of wave number. Similar to what was observed for the transverse modes, the imaginary part of the longitudinal modes becomes constant at large wave number. The real parts of the longitudinal modes do not show any qualitative variation at large wave number as we change the sign of the coupling term.

\begin{figure}[ht]
\begin{center}
\includegraphics[scale=0.35]{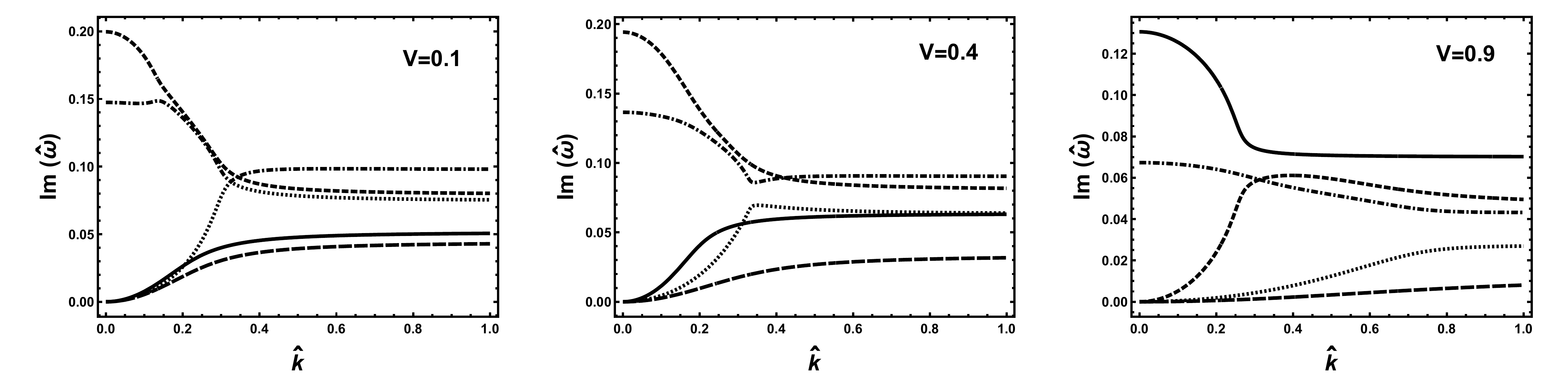}\\
\includegraphics[scale=0.35]{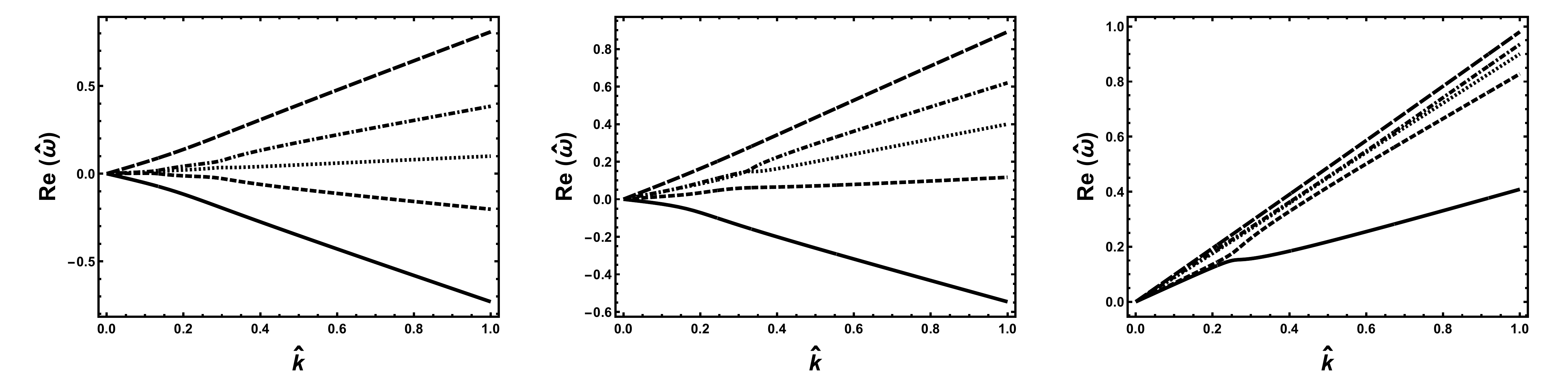}
\caption{Real and imaginary parts of the longitudinal modes considering a negative value for the product of the coupling terms, $\mathcal{\hat{L}}_{\pi
n}\mathcal{\hat{L}}_{n\pi }=-1$, for three different values of background velocity, $V=0.1$, $V=0.4$, and $V=0.9$.}
\label{longrestcoup}
\end{center}
\end{figure}

\begin{figure}[ht]
\begin{center}
\includegraphics[scale=0.35]{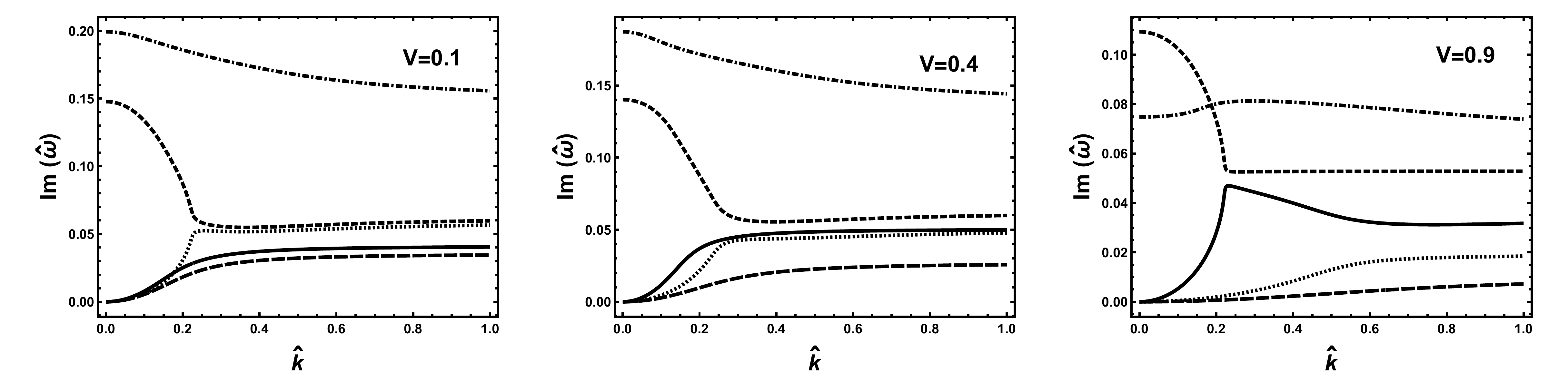}\\
\includegraphics[scale=0.35]{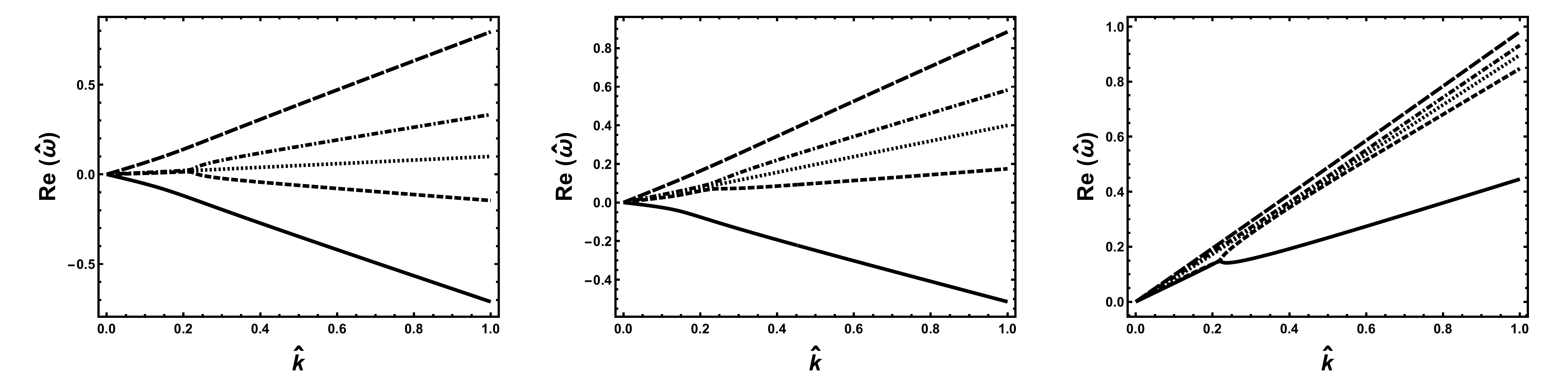}
\caption{Real and imaginary parts of the longitudinal modes considering a positive value for the product of the coupling term, $\mathcal{\hat{L}}_{\pi n}\mathcal{\hat{L}}_{n\pi }=2$, for three different values of background velocity $V=0.1$, $V=0.4$, and $V=0.9$.}
\label{longrestcoup2}
\end{center}
\end{figure}

Now we discuss perturbations on top of a moving background fluid. Again, we assume that the 4-velocity and the perturbations are taken in the same direction, e.g., the $x$ axis. Therefore, we have $u_{0}^{\mu }=\gamma(1,V,0,0)$ and $k^{\mu}=(\omega,k,0,0)$, which leads once more to the expressions of $\hat\Omega$ and $\hat\kappa$ given in Eqs.~(\ref{omega}) and (\ref{kappa}). In Figs.~\ref{longrestcoup} and \ref{longrestcoup2}, we display the solutions of the dispersion relation in Eq.~(\ref{disper04}), substituting Eqs.~(\ref{omega}) and (\ref{kappa}), considering several values of velocity and product of the coupling terms. In the following, we shall study the properties of the solutions at vanishing wave number, $\hat{k}=0$, in order to extract a set of necessary linear stability conditions, as already done in the previous sections. In this case, the dispersion relation reads
\begin{equation}
\left( \gamma \hat{\omega}\right) ^{3}\left[ -3\mathcal{A}\left( 1-2\mathcal{%
S}V^{2}+\mathcal{M}V^{4}\right) \gamma ^{2}\hat{\omega}^{2}+i\left( \hat{\tau%
}_{\kappa }V^{4}+3\mathcal{B}-\mathcal{B}V^{2}\right) \gamma \hat{\omega}%
+3-V^{2}\right] =0,
\end{equation}%
with two nonvanishing solutions, corresponding to the two longitudinal nonhydrodynamic
modes of the theory, 
\begin{equation}
\gamma \hat{\omega}_{\pm}=i\frac{\hat{\tau}_{\kappa }V^{4}+\mathcal{B}\left(
3-\mathcal{D}V^{2}\right) \pm \sqrt{\left[ \hat{\tau}_{\kappa }V^{4}+\mathcal{B}\left(
3-\mathcal{D}V^{2}\right) \right] ^{2}-12\mathcal{A}\left( 3-V^{2}\right) \left( 1-2%
\mathcal{S}V^{2}+\mathcal{M}V^{4}\right) }}{3\mathcal{A}\left( 1-2\mathcal{S}%
V^{2}+\mathcal{M}V^{4}\right) }. \label{longeq_final}
\end{equation}

As already discussed, in order to guarantee the stability of these modes, we must have that the imaginary part of the dispersion relation is positive for all possible values of the velocity. Evidently, this is achieved if both numerator and denominator carry the same sign. The denominator in Eq.~(\ref{longeq_final}) is positive for $V=0$, and thus it must remain positive for all values of $V$ -- a change of sign would already imply an instability. Therefore, we must have
\begin{equation}
1-2\mathcal{S}V^{2}+\mathcal{M}V^{4}>0,\text{  }\forall\hspace{0.3cm} 0\leq V\leq1.\label{denom}
\end{equation}
Here, Eq.~(\ref{denom}) is a polynomial function which is quadratic in $V^{2}$ and positive at $V=0$. The condition above is satisfied if the smallest root of this polynomial is larger than 1. This guarantees that the function is positive in the physical interval of $0\leq V^{2}\leq 1$ and instabilities would only occur in the nonphysical region in which the background velocity is greater than the speed of light. This is ensured by the following inequality

\begin{equation}
\mathcal{S}-\mathcal{R}\geq \mathcal{M}.\label{s_r}
\end{equation}
Note that this inequality leads to the causality condition derived in Eq.~(\ref{causality_coup_long}) for perturbations on top of a fluid at rest. This can be seen by taking the definition of $\mathcal{M}=\mathcal{S}^2-\mathcal{R}^2$ and then
\begin{equation}
\mathcal{S}-\mathcal{R}\geq(\mathcal{S}-\mathcal{R})(\mathcal{S}+\mathcal{R})\Longrightarrow\mathcal{S}+\mathcal{R}\leq1\Rightarrow -\frac{3}{2}\hat{\tau}_{\pi }\left( \hat{\tau%
}_{n}-\hat{\tau}_{\kappa }\right) -\hat{\tau}_{n}\left( \hat{\tau}_{\pi
}-2\right) \leq \mathcal{\hat{L}}_{\pi n}\mathcal{\hat{L}}_{n\pi },\label{cond1}
\end{equation}
where we used the previously derived stability conditions $\mathcal{S}-\mathcal{R}>0$ and $\mathcal{R}\geq0$. Once again, we see that the stability conditions obtained for perturbations on top of a moving background are related to the causality conditions satisfied by perturbations on top of a static background. The relations in Eq.~(\ref{cond1}) can be used to derive the additional relations
\begin{eqnarray}
\mathcal{M}&=&(\mathcal{S}+\mathcal{R})(\mathcal{S}-\mathcal{R})\leq1\Rightarrow \frac{3}{2}\hat{\tau}_{\kappa }\left( \hat{%
\tau}_{\pi }+4\right) -\frac{9}{2}\mathcal{A}\leq \mathcal{\hat{L}}_{\pi n}%
\mathcal{\hat{L}}_{n\pi },\label{cond2}\\
\mathcal{S}&\leq&\frac{1+\mathcal{M}}{2}\Rightarrow -\frac{2}{3}\mathcal{%
\hat{L}}_{\pi n}\mathcal{\hat{L}}_{n\pi }\leq \left( \hat{\tau}_{\pi
}-2\right) \left( \hat{\tau}_{n}-\hat{\tau}_{\kappa }\right).\label{cond3}
\end{eqnarray}

In order to have stable modes, the numerator in Eq.~(\ref{longeq_final}) must also be positive definite. One can show that the term inside the square root in Eq.~(\ref{longeq_final}) is positive definite as long as we assume that the product of the coupling terms is negative. Therefore, in order for the modes $\hat{\omega}_+$ and $\hat{\omega}_-$ to have a positive definite imaginary part it is sufficient to impose that
\begin{equation}
\frac{\hat{\tau}_{\kappa }}{3\mathcal{B}}V^{4}-\frac{\mathcal{D}}{3}V^{2}+1
\geq0,\text{  }\forall\hspace{0.3cm} 0\leq V\leq1.
\end{equation}
In order for the inequality to be fulfilled, the smallest root of the polynomial above must be greater than 1, rendering the function positive in the physical interval in which the background velocity does not exceed the speed of light. The analysis here is analogous to the one performed on Eq.~(\ref{denom}), leading to a new constraint for the relaxation times
\begin{equation}
\hat{\tau}_{\pi}+\hat{\tau}_{n}\geq\hat{\tau}_{\kappa}+2,
\end{equation}
which is clearly satisfied for by the transport coefficients calculated from the Boltzmann equation \cite{is1, dnmr, 14moment} and used in the previous section of this paper. 

Assuming the coupling terms satisfy $\mathcal{\hat{L}}_{\pi n}\mathcal{\hat{L}}_{n\pi }<0$, the stability conditions derived for the longitudinal modes can be summarized as
\begin{equation}
\hat{\tau}_{\pi }\geq 2,\left. {}\right. \hat{\tau}_{n}\geq \hat{\tau}%
_{\kappa },\left. {}\right. |\mathcal{\hat{L}}_{\pi n}\mathcal{%
\hat{L}}_{n\pi }|\leq \frac{3}{2}\left( \hat{\tau}_{\pi
}-2\right) \left( \hat{\tau}_{n}-\hat{\tau}_{\kappa }\right). \label{conditions_final}
\end{equation}
From the transverse modes, we had the additional condition%
\begin{equation}
|\mathcal{\hat{L}}_{\pi n}\mathcal{\hat{L}}_{n\pi }|
\leq 2\hat{\tau}_{n}\left( \hat{\tau}_{\pi }-1\right) .
\end{equation}
However, one can demonstrate that this is contained in the inequalities in Eq.~(\ref{conditions_final}). Therefore, the conditions for $\hat{\tau}_{\pi }$ and $\hat{\tau}_{n}$ are the same as the ones obtained without coupling terms and we just obtain an additional constraint for $\mathcal{\hat{L}}_{\pi n}\mathcal{\hat{L}}_{n\pi }$. If we use the relaxation times
calculated from the Boltzmann equation,
\begin{equation}
|\mathcal{\hat{L}}_{\pi n}\mathcal{\hat{L}}_{n\pi }|\leq\frac{891}{32}\approx27.84. \label{stab_cond2}
\end{equation}

\begin{figure}[ht]
\begin{center}
\includegraphics[scale=0.45]{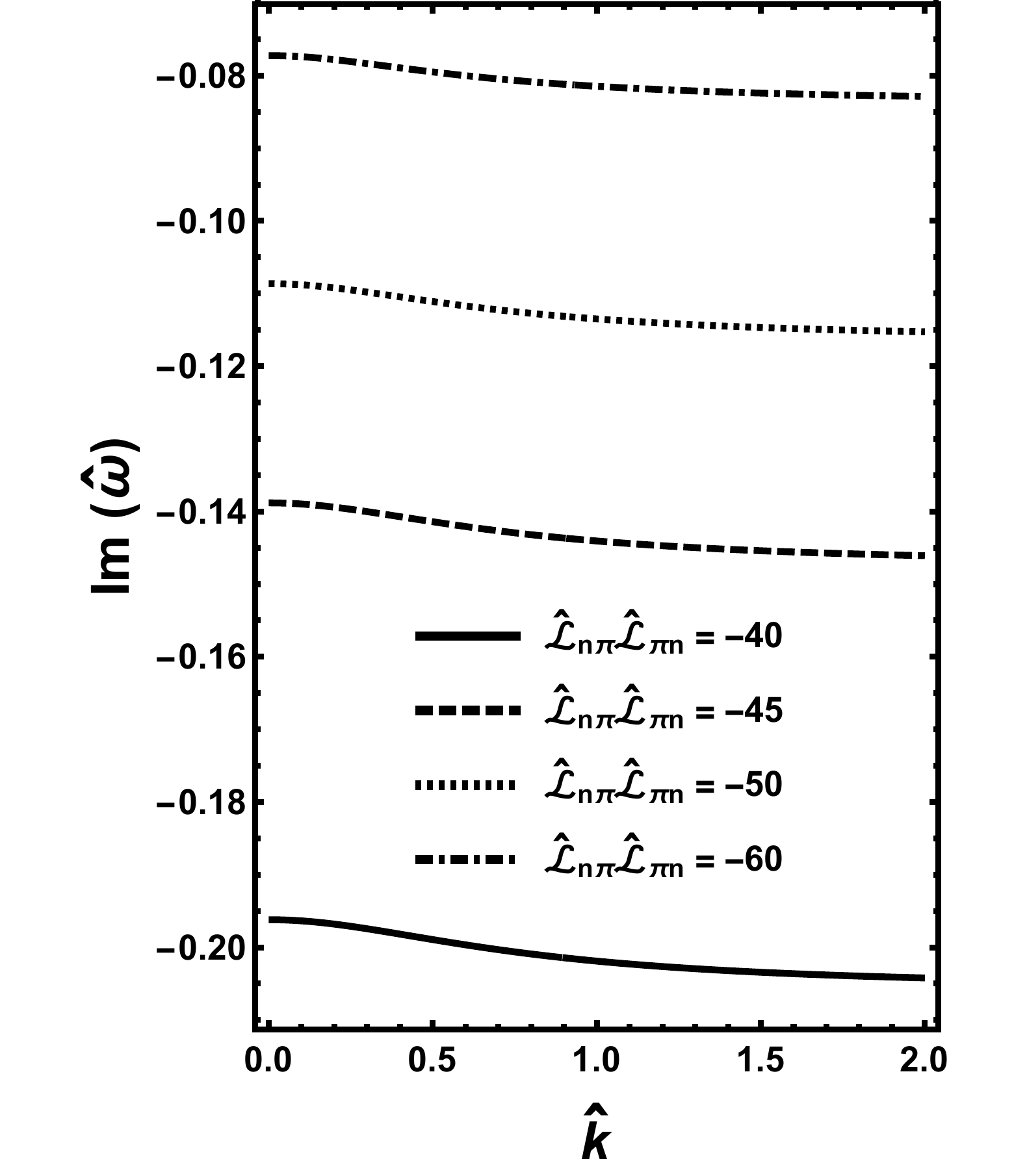}
\caption{Imaginary part of the unstable shear modes for different negative values of the coupling term, $\mathcal{\hat{L}}_{\pi
n}\mathcal{\hat{L}}_{n\pi }=-40, -45, -50, -60$, for $V=0.99$.}
\label{longrestcoup3}
\end{center}
\end{figure}

For the sake of completeness, the solution of Eq.~(\ref{disper04}) for perturbations of top of a moving background are displayed in Figs.~\ref{longrestcoup} and \ref{longrestcoup2} considering a negative and a positive value for the product of the coupling terms,  $\mathcal{\hat{L}}_{\pi n}\mathcal{\hat{L}}_{n\pi }=-1$ and $\mathcal{\hat{L}}_{\pi n}\mathcal{\hat{L}}_{n\pi }=2$, respectively. We also show several examples of configurations that are driven unstable by the coupling terms in Fig.~\ref{longrestcoup3}, taking the following values for the product of the coupling terms that violate Eq.~(\ref{stab_cond2}), $\mathcal{\hat{L}}_{\pi
n}\mathcal{\hat{L}}_{n\pi }=-40, -45, -50, -60$.

\section{Conclusions and Remarks}
\label{conc}

In this work we presented a linear stability analysis of Israel-Stewart theory including the effects of shear-stress tensor and net-baryon diffusion current. In particular, we investigated the effects that second-order terms (that are linear in the dissipative currents) that couple one dissipative current with the other, the diffusion-viscous couplings, can have on the stability and causality of the theory. In order to achieve this goal, we considered small perturbations around a global equilibrium state with energy density $\varepsilon_0$, vanishing net-baryon number, and a finite fluid 4-velocity $u^\mu_0$. The stability condition is that the system returns to this global equilibrium state after being perturbed.

We first considered the case where the coupling terms are zero. In this case, the modes related to fluctuations of energy, momentum, and net-baryon number decouple. The dispersion relation for the modes related to fluctuations of energy and momentum obtained here are identical to the ones first derived in Ref.~\cite{rischke}. We thus obtained the same stability conditions for the shear relaxation time, $\tau_\pi$, as in Ref.~\cite{rischke}, given by 
\begin{equation*}
\tau_\pi\geq\frac{2\eta}{\varepsilon_0+P_0},
\end{equation*}
where $\eta$ is the shear viscosity, $\varepsilon_0$ is the energy density, and $P_0$ is the thermodynamic pressure. Furthermore, since we considered the effects of net-baryon current, we also obtained a stability condition for the net-baryon diffusion relaxation time, $\tau_n$,
\begin{equation*}
    \tau_n\geq\frac{\kappa_n}{\bar{n}_B},
\end{equation*}
where $\kappa_n$ is the diffusion coefficient and $\bar{n}_{B}$ the baryon number density.

We then investigated how the introduction of the aforementioned coupling terms affects these stability conditions. In this scenario, we proved that the theory is not stable for arbitrary values of such coupling terms. In order to be consistent with kinetic theory calculations and the derivation of fluid dynamics from the second law of thermodynamics, we assumed that the product of the coupling terms is negative. We then showed that the linear stability conditions for the relaxation times derived in the presence of diffusion-viscous coupling are not modified by the inclusion of the coupling terms, and remain being the ones listed above. Furthermore, we obtained a stability condition that must be satisfied by the coupling terms themselves, $\ell_{\pi n}$ and $\ell_{n\pi }$, given by
\begin{equation*}
|\ell_{\pi n}\ell_{n\pi }|\leq \frac{3}{2}\left(\tau_{\pi
}-\frac{2\eta}{\varepsilon_0+P_0}\right) \left(\tau_{n}-\frac{\kappa_n}{\bar{n}_B}\right).
\end{equation*}
Once more, we emphasized that these conditions are obtained assuming $\ell_{\pi n}\ell_{n\pi }<0$. However, we showed some examples that the system can be stable for $\ell_{\pi n}\ell_{n\pi }>0$, but these cases were not studied thoroughly.

We confirmed in this paper that the linear causality conditions obtained for perturbations around a background fluid at rest are equivalent to linear stability conditions for perturbations on top of a moving fluid, as it was first derived in Ref.~\cite{rischke}. Finally, we note that the stability conditions obtained by Olson in Ref.~\cite{olson} also include the effects of diffusion-viscous coupling and even bulk viscous pressure, which was neglected here. However, in this case the constraints for the transport coefficients were written in a more convoluted form (constraints for the diffusion-viscous coupling terms were not explicitly derived). Furthermore, and more importantly, the linear stability conditions obtained by Olson cannot be trivially extended to the case of a vanishing background net-charge density and, thus, it is not possible to directly compare them to our results. This happens because the perturbations defined by Olson diverge in the limit of vanishing net charge, which is inconsistent with the assumptions made by Olson when deriving the linear stability conditions (in Theorem A of Ref.~\cite{olson}, Olson assumes that the perturbations do not diverge when deriving linear stability conditions).

Nevertheless, some of the linear \textit{causality} conditions derived by Olson are equivalent to those derived in this paper. The linear causality conditions for the transverse characteristic velocities calculated by Olson [Eq.~(91) of Ref.~\cite{olson}] are equivalent to the corresponding linear causality conditions derived in this paper, see Eq.~(\ref{causal_coup_trans}), in the limit of vanishing net charge. The same does not occur when comparing the linear causality conditions derived for the longitudinal modes. In this case, the difficulty in the comparison lies in taking the limit of vanishing bulk viscosity and relaxation time, for which case the result diverges. In order to compare with Olson's result, we would need to include the effects of bulk viscosity in our calculation as well. However, we leave this task for future work.

\section*{Acknowledgements}

The authors thank J.~Noronha for insightful discussions. G.S.D. thanks Conselho Nacional de Desenvolvimento Cient\'ifico e Tecnol\'ogico for support and Funda\c c\~ao Carlos Chagas Filho de Amparo \`a Pesquisa do Estado do Rio de Janeiro (FAPERJ), Grant No. E-26/202.747/2018. C.V.B. thanks Coordena\c c\~ao de Aperfei\c coamento de Pessoal de N\' ivel Superior and FAPERJ for support.

\bibliographystyle{apsrev4-1}
\bibliography{refs}

\end{document}